%% file: main.tex
\documentclass[10pt,nocopyrightspace]{sigplanconf}
\pdfoutput=1

\usepackage[normalem]{ulem}
\usepackage{epsfig,endnotes}
\usepackage{xspace}
\usepackage{hyperref}
\usepackage{color}
\usepackage{tabularx}
\usepackage{booktabs}
\usepackage{graphicx}
\usepackage{amsmath}
\usepackage{pifont}
\usepackage{natbib}
\usepackage{comment}
\usepackage{url}
\usepackage{enumitem}
\usepackage{multirow}

\makeatletter
\g@addto@macro{\UrlBreaks}{\UrlOrds}
\makeatother

\newcommand{\cmark}{\ding{51}}

\newcommand{\point}[1]{\par\smallskip\noindent{\bf #1}}
\newcommand{\bench}[1]{\textbf{\small #1}}
\newcommand{\niceunitk}{\,{\small K}\xspace}
\newcommand{\niceunitkb}{\,{\small KB}\xspace}
\newcommand{\niceunitmb}{\,{\small MB}\xspace}
\newcommand{\niceunits}{\,{\small s}\xspace}
\newcommand{\niceunitkbs}{\,{\small $\text{KB}/\text{s}$}\xspace}
\newcommand{\niceunitpct}{\%\xspace}

\newcommand{\smtexttt}[1]{{\small{\texttt{#1}}}}

  {\begin{list}{$\bullet$}%
     {\setlength{\parsep}{0pt}%
      \setlength{\topsep}{0pt}%
      \setlength{\itemsep}{0pt}}}%
  {\end{list}}

\newcommand{\sys}{\textsc{McFly}\xspace}

\definecolor{lightgray}{rgb}{.9,.9,.9}

{{\ttfamily \hyphenchar\the\font=`\-}

\bibliographystyle{ACM-Reference-Format}

\begin{document}
%don't want date printed
%\date{}

\title{\sys: Time-Travel Debugging for the Web}

\authorinfo{John~Vilk, Emery~D.~Berger}{University of Massachusetts Amherst}{\{jvilk, emery\}@cs.umass.edu}
\authorinfo{James~Mickens}{Harvard University}{mickens@g.harvard.edu}
\authorinfo{Mark~Marron}{Microsoft Research}{marron@microsoft.com}

%\author{John Vilk}
%\affiliation{
%  \department{College of Information and Computer Sciences}
%  \institution{University of Massachusetts Amherst}
%  \country{USA}
%}
%\email{jvilk@cs.umass.edu}

%\author{Emery~D.~Berger}
%\affiliation{
%  \department{College of Information and Computer Sciences}
%  \institution{University of Massachusetts Amherst}
%  \country{USA}
%}
%\email{emery@cs.umass.edu}

%\author{James Mickens}
%\affiliation{
%  \department{Department of Computer Science}
%  \institution{Harvard University}
%  \country{USA}
%}
%\email{mickens@g.harvard.edu}

%\author{Mark Marron}
%\affiliation{
%  \department{Microsoft Research}
%  \institution{Microsoft Corporation}
%  \country{USA}
%}
%\email{marron@microsoft.com}

\date{}
\maketitle

\thispagestyle{empty}

\input{abstract}
\input{introduction}

\input{background-mr-ttd}

\input{time-travel}

\input{deterministic}

\input{evaluation}

%\input{discussion}

\input{related-work}

\input{conclusion}

\input{acknowledgements}

\clearpage
{
  %\footnotesize
\bibliographystyle{plain}
\bibliography{main}
}

\end{document}

%% file: abstract.tex
% !TeX root = main.tex
\begin{abstract}

Time-traveling debuggers offer the promise of
simplifying debugging by letting developers freely step forwards
\emph{and backwards} through a program's execution.  However, web
applications present multiple challenges that make time-travel
debugging especially difficult. A time-traveling debugger for web
applications must accurately reproduce all network interactions,
asynchronous events, and visual states observed during the original
execution, both while stepping forwards and backwards.  This must all
be done in the context of a complex and highly multithreaded browser
runtime. At the same time, to be practical, a time-traveling debugger must maintain
interactive speeds.

This paper presents \sys, the first time-traveling debugger for web
applications. \sys departs from previous approaches by operating on a
high-level representation of the browser's internal state. This
approach lets \sys provide accurate time-travel
debugging---maintaining JavaScript and visual state in sync at all
times--at interactive speeds. \sys's architecture is browser-agnostic,
building on web standards supported by all major browsers.  We
have implemented \sys as an extension to the Microsoft Edge web browser, and
core parts of \sys have been integrated into a time-traveling debugger
product from Microsoft.
\end{abstract}

%% file: introduction.tex
\section{Introduction}
\label{sec:intro}

Web applications are notoriously frustrating to debug.
JavaScript events can race with one another, leading to Heisenbugs.
Network requests can intermittently fail or return unexpected results, leading to unanticipated error states.
JavaScript code can also race with the browser itself via the document object model (DOM), since independent browser threads update visual state in parallel with JavaScript execution.
These behaviors can make application bugs challenging for developers to diagnose and fix.

To aid debugging, all modern web browsers include integrated debuggers~\cite{edge-debugger,chrome-debugger,firefox-debugger,safari-debugger}.
While these can be useful, they often provide little assistance to developers.
If a bug is the result of a specific event order (a Heisenbug), the act of debugging can disrupt the event schedule and prevent the bug from appearing.
Even when a bug does recur while debugging, identifying its root cause can be difficult: in event-driven settings like the web, bug symptoms can manifest far from their root causes.

There are two widely used approaches to find these bugs: scattering logging statements around the program (a.k.a. ``printf debugging'' using \smtexttt{console.log}), or placing breakpoints to pause the program at specific statements.
Both are laborious and iterative processes.
Poring over logs to identify bugs often reveals the need to rerun the program with new logging statements in place.
Using breakpoints, developers step the program forwards until the first sign that something has gone wrong.
Unfortunately, if the breakpoint does not precede the root cause of the bug, the developer must reset breakpoints and restart execution from the beginning.
As a result, debugging is currently an arduous and painstaking process for web developers.

A widely-proposed debugging approach intended to address these difficulties is \emph{reversible} or \emph{time-travel debugging}.
A time-traveling debugger would let developers step forwards and backwards through the execution.
Rather than needing to repeatedly restart execution from the beginning, placing breakpoints, or adding logging, developers would be able to step forwards until the first symptom of a bug appears, and then step backwards to isolate its root cause.
Developers report that they step too far forward during debugging sessions ``all of the time'', and that a ``back in time [i.e., time-traveling] debugger would be wonderful''~\cite{debugger-quotes}.

% TODO: Section pointers in diagram for where we actually support these items?
\begin{figure}
\centering\includegraphics[width=0.48\textwidth]{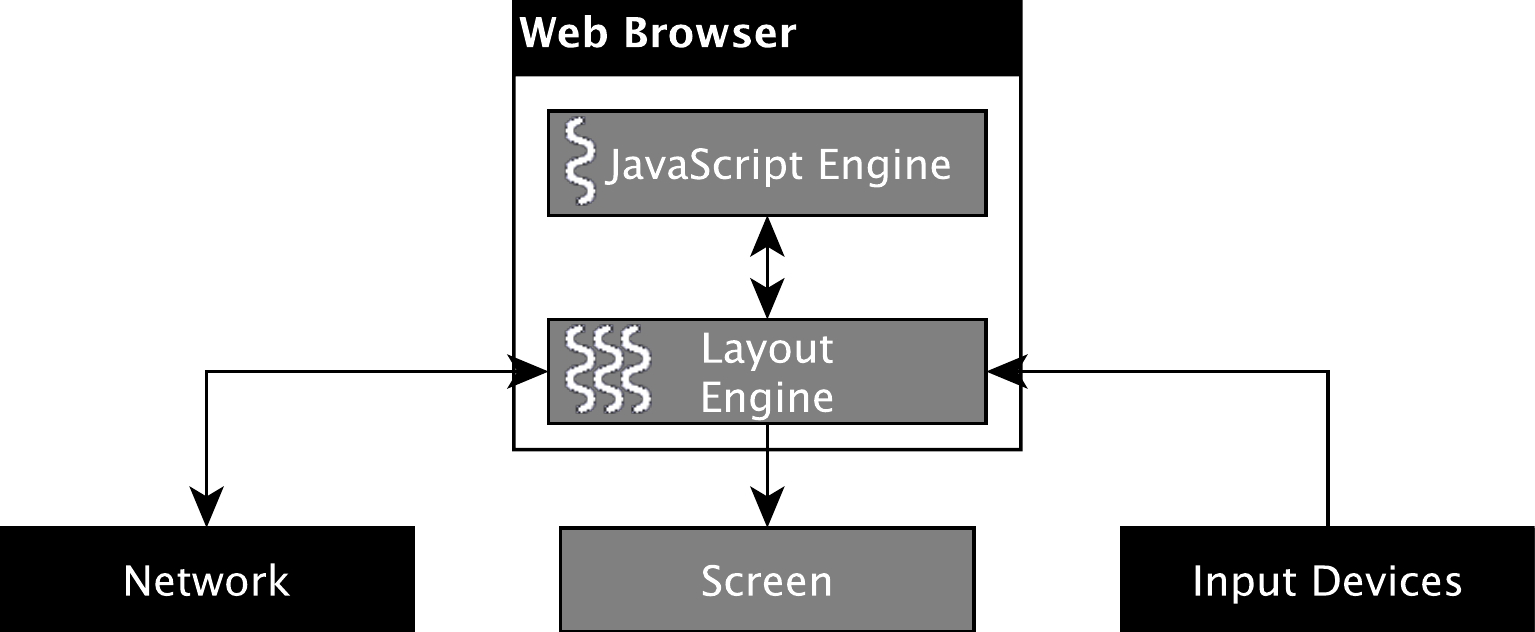}
	\caption{\textbf{All major web browsers separate web application computation (JavaScript engine) from I/O (Layout Engine).} \sys both synchronizes and maintains visibility into the items shown in gray (the JavaScript engine, the multithreaded layout engine, and the screen), all at interactive speed.}
	\label{fig:browser-arch}	
\end{figure}

%Developers have long sought after time-traveling debuggers (TTDs) because they promise to ease error diagnosis by letting developers freely step forward and backward through a program execution.
%TTDs provide reverse complements to existing stepping debugger features, such as \emph{step backward} complementing \emph{step forward}.
%With a TTD, developers can reproduce a bug once and then work backwards from a bug's symptom to isolate its root cause.
%The TTD ensures that the developer observes the same program states during time-travel as the original execution. 

% Move checkboxes down to related work.
% Could say: McFly Combines previous approaches, builds on them? to simultaneously meet all of these challenges. Incorporates prev. approaches and builds on them. Timelapse, the step-backward stuff, tardis, etc.
% Need to discuss challenges. Abstract mentions reproducing states in context of complex + multithreaded browser runtime.
Time-travel debugging has been extensively explored in non-web-based settings~\cite{tardis, kingVM, boothe, undodb, burtsev, revirt, dunlap-revirt-2, retrace, chronon, omniscientTOD, odb, tralfamadore, bttf, pothier-sod, koju05}.
In some of these settings, time-travel debugging is essentially a solved problem.
% This is understood in simple contexts; "solved problem".
% Logging everything or periodic checkpoints.
For example, in serial computations with no side effects and no I/O, it is always possible to replay the application from the beginning to the desired point in time.
As a standard optimization, time-traveling debuggers take periodic checkpoints of program state (e.g., mutable data and the program counter) and resume execution from the nearest checkpoint.
Replay can further be accelerated by eliding pauses resulting from user and network interactions.
To address side-effects and I/O, they can log operations during the first execution and replay them during time-travel.

Unfortunately, \emph{past approaches are not sufficient to enable time-travel debugging for web applications.}
Unlike traditional environments, web applications are deeply entangled with their visual state and cannot be debugged in isolation.
Most web application bugs are related to interactions with the layout engine~\cite{frolin-bugs}, and many manifest visually.
Because there is no way to checkpoint and roll back a web application's visual state, a time-traveling debugger would need to re-execute the program from the beginning of time \emph{every time the developer steps backwards}.
Even an optimized re-execution that elides pauses will eventually become too slow to be practical.
Given that a web application's visual state is maintained within a complex multithreaded browser, it is not immediately obvious how a time-traveling debugger could efficiently support visual state.

This paper presents \sys, the first time-traveling debugger for web applications.\footnote{A video from 2015 shows \sys in action: \url{https://channel9.msdn.com/blogs/Marron/Time-Travel-Debugging-for-JavaScriptHTML}}
\sys overcomes the challenges of the web environment by capturing not only JavaScript program state but also a high-level representation of the layout engine's internal state, enabling visual state checkpointing and replay.
% Verified that this is true in Firefox:
% https://dxr.mozilla.org/mozilla-central/source/layout/style/nsAnimationManager.h#123
%Chrome (uses a Timer class that has an iteration count):
% https://cs.chromium.org/chromium/src/third_party/WebKit/Source/core/animation/css/CSSAnimations.h?sq=package:chromium&dr=Ss&l=115
For example, a CSS animation could easily modify thousands of pixels of on-screen content per second.
Instead of tracking these pixels, \sys captures the internal frame counter, which represents the animation in a single integer.
\sys captures high-level state for all core layout engine functionality, including DOM elements, event listeners, and network connections.
Because this high-level state is formally described in web standards documents~\cite{domSpec}, \sys's approach is browser-agnostic.
During debugging, \sys manages this high-level state to keep the JavaScript engine and the multithreaded layout engine in sync, making it possible for developers to efficiently step backwards.

We have implemented a prototype version of \sys as an extension to the Microsoft Edge web browser, and show that its approach enables interactive debugging speeds.
Core parts of \sys have been integrated into a time-traveling debugger product from Microsoft~\cite{chakra-ttd}.

\subsection*{Contributions}

This paper makes the following contributions:

\begin{itemize}[itemsep=1pt,topsep=2pt,leftmargin=10pt]
  \item It details the challenges to time-travel debugging web applications~(\S\ref{sec:background}).
  \item It proposes a browser-agnostic architecture for a time-traveling debugger for web applications~(\S\ref{sec:sys}).
  \item It describes \sys, a prototype time-traveling debugger for web applications (\S\ref{sec:implementation}) that can achieve interactive debugging speeds~(\S\ref{sec:eval}).
\end{itemize}

%\begin{figure}
%	\centering\includegraphics[width=0.45\textwidth]{images/breakpoint-debug.png}
%	\caption{Standard browser debuggers, such as Microsoft Edge's developer tools, let developers inspect DOM state and view the application's GUI at breakpoints. Ignoring GUI state during time-travel prevents these tools from working.}
%	\label{fig:breakpoint-debug}	
%\end{figure}

%% file: background-mr-ttd.tex
% !TeX root = main.tex
\section{Challenges to Time-Travel Debugging Web Applications}
\label{sec:background}

% Challenges:
% - Must capture all sources of nondet; there are many.
% - Cannot modify web browser in intrusive way. Large and complicated, perf, etc.
% - Internal browser concurrency???

Time-travel debugging for web applications is challenging for a variety of reasons.
To be \emph{useful}, a time-traveling debugger must provide the developer with visibility into the application's visual state (\S\ref{subsec:vis-state}).
In particular, it must keep the JavaScript engine, the layout engine, and the screen in sync at all times (\autoref{fig:browser-arch}).
To ensure that time-travel is \emph{deterministic}, it must correctly capture all I/O and control for nondeterminism (\S\ref{subsec:nondet}).
To be \emph{usable}, it must be able to checkpoint and roll back the application's visual state (\S\ref{subsec:ckpt-vis-state}), enabling interactive speeds. 
Finally, to be \emph{practical}, it must be straightforward to integrate into modern browsers, which are complex multithreaded applications consisting of millions of lines of code (\S\ref{subsec:portability}).

\subsection{Supporting Visual State}
\label{subsec:vis-state}

\begin{figure}[tb]
	\centering\includegraphics[width=0.48\textwidth]{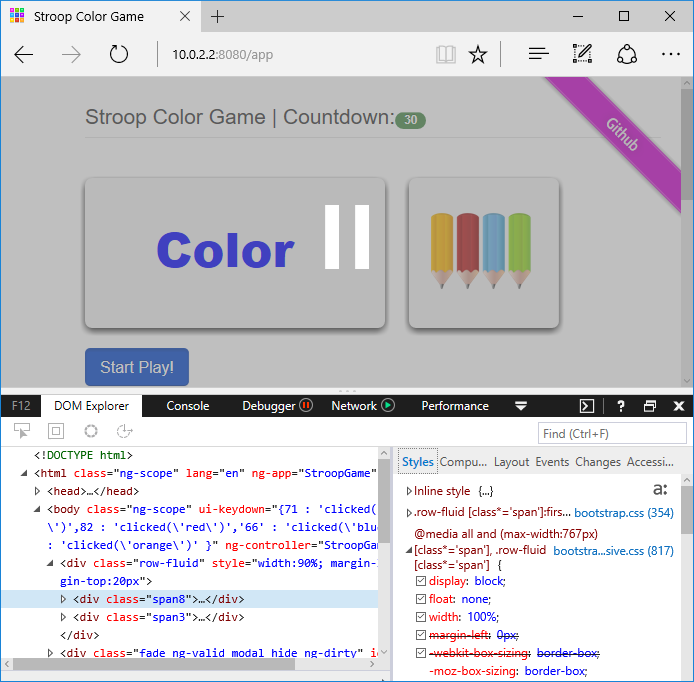}
	\caption{\textbf{A web application's visual state is critical for debugging.} Existing browser debugging tools, such as Microsoft Edge's DOM Explorer shown above, let developers examine this state.
	A time-traveling debugger that ignores visual state would prevent these tools from working, and would display the application as a blank screen.
	\sys supports debugging tools like the DOM Explorer because it ensures that visual state remains in sync with JavaScript execution.}
	\label{fig:visual-debug}	
\end{figure}

Debugging web applications is often a visual experience; when an application does not produce the correct visual output, the developer needs to figure out why.
Browsers contain debugging tools, like Microsoft Edge's DOM Explorer (\autoref{fig:visual-debug}), that let developers examine this state and map it back to the pixels displayed on the screen.
Using the DOM Explorer, developers can determine which CSS and HTML properties influence a UI element's appearance, what JavaScript code is listening for input events, and where a UI element appears on the screen.

To be useful, a time-traveling debugger must support visual state during stepping operations.
Failing to do so would make debugging vastly more difficult; developers would not be able to see the GUI or examine its underlying properties in the DOM Explorer.

However, this visual state is not managed by the JavaScript engine; instead, it is managed within the layout engine.
Every major web browser has its own layout engine; Blink (for Chrome)~\cite{blink}, WebKit (for Safari)~\cite{webkit}, Gecko (for Firefox)~\cite{gecko}, and EdgeHTML (for Microsoft Edge)~\cite{edgehtml}.
JavaScript code interacts with the layout engine via a set of standard interfaces commonly referred to as the Document Object Model (DOM)~\cite{domSpec}.

In order to support visual state during debugging sessions, a time-traveling debugger needs to keep the layout engine and the screen in sync with JavaScript execution.
However, in all major web browsers, many visual updates occur within the layout engine and independent of JavaScript execution.
Network dependencies like images begin downloading as soon as they are inserted in the UI and appear as soon as they finish loading.
CSS animations manipulate GUI elements at 60 frames per second in parallel with JavaScript execution.
The debugger must somehow manage these updates to ensure that visual state remains synchronized with the debugging session at all times.

\subsection{Capturing I/O and Controlling for Nondeterminism}
\label{subsec:nondet}

A time-travel debugging session can diverge from the original execution if the debugger fails to reproduce the same I/O and nondeterminism observed in the original execution.
This divergence can change the control flow of the program and disrupt stepping operations.
Thus, it is crucial that a time-traveling debugger accurately capture I/O and control for nondeterminism to ensure a deterministic debugging experience.

To capture I/O, the debugger needs to operate within the layout engine, as I/O operations can be triggered implicitly and explicitly by HTML, CSS, and JavaScript.
For example, HTML and CSS can reference external images, which the browser loads asynchronously and independent of JavaScript execution.
Via the DOM, JavaScript code can send HTTP requests and process HTTP responses.
All of these I/O operations must be captured and synchronized with JavaScript execution during debugging sessions.

The debugger must manage several sources of nondeterminism, including random numbers, JavaScript event orderings, and thread interleavings within the multithreaded layout engine.
While random number generation can easily be made deterministic, the remaining two sources of nondeterminism are more complicated to manage.

The order of JavaScript events ultimately determines the high-level control flow of a web application because JavaScript execution is completely event-driven and occurs on a single hardware thread.
%A time-traveling debugger must reproduce the same JavaScript events in the same order as the original execution.
The layout engine maintains an event queue that contains events that are waiting to execute.
JavaScript code subscribes to specific events, such as mouse clicks on a button, using JavaScript event listeners.
The layout engine dispatches events to the event queue when they occur, which eventually run the relevant event listeners.
During stepping operations, a time-traveling debugger must ensure that events exit the event queue in the same order as the original execution.

Thread interleavings within the multithreaded layout engine determine the order in which certain visual state updates occur.
In particular, all major layout engines use a dedicated thread for CSS animations, which manipulates GUI items in parallel with JavaScript execution in order to maintain consistent frame rates.
Some major layout engines also update the GUI with asynchronously loaded network resources in parallel with JavaScript execution.
A time-traveling debugger must ensure that these visual state updates are synchronized with JavaScript execution at all times.
Failing to reproduce these (benign) data races can result in program replay that diverges from the original execution, which prevents time-travel operations from completing successfully.

\subsection{Checkpointing Visual State}
\label{subsec:ckpt-vis-state}

A time-traveling debugger must be able to checkpoint and roll back a web application's visual state in order to support stepping operations at interactive speeds.
Without these checkpoints, a time-traveling debugger would need to deterministically replay the application from the beginning to support stepping operations, which would be unusably slow on long traces.

The layout engine manages visual state and determines how this state translates into pixels displayed on the screen.
A time-traveling debugger must somehow capture this visual state into a checkpoint, and ensure that rolling back to the checkpoint results in the same screen pixels.
However, visual state does not exist independent of JavaScript program state; instead, the two are deeply intertwined.
JavaScript objects can retain visual state, such as UI elements that are not visible on the screen.
UI elements appear to JavaScript code as ordinary JavaScript objects, but internally, they reflect visual state stored within the layout engine.
The layout engine can retain program state, as it stores references to JavaScript functions that have been registered as event listeners.
Thus, a time-traveling debugger must accurately checkpoint and restore dependencies between visual and program state.

\subsection{Portability Across Browsers}
\label{subsec:portability}

Modern web browsers are complex pieces of software with millions of lines of code~\cite{browsers-loc}.
These browsers are aggressively optimized to edge out the competition on benchmarks.
It would be infeasible to significantly change how the browser operates in order to support time-travel debugging.
Instead, to be practical, changes must be unintrusive, maintainable, and portable across browsers, which places severe restrictions on how time-travel can be implemented.

%% file: time-travel.tex
% !TeX root = main.tex
% TODO: Motivate replay strategies. When do you use each?
\section{\sys}
\label{sec:sys}

\sys is our prototype time-traveling debugger for web applications that overcomes the challenges presented in \autoref{sec:background}.
\sys operates on a high-level representation of a browser's internal state, letting it provide accurate time-travel debugging with support for visual state at interactive speeds. 
In this section, we present \sys's architecture, which is browser-agnostic.

\subsection{Time-Travel Overview}
\label{subsec:tt-overview}

We first provide an overview of how a developer uses \sys to debug a web application, which structures the remainder of this section.

\point{Reproducing a bug:}
The developer loads the web application with \sys open, and interacts with the application until they discover buggy behavior.
While this happens, \sys interacts with the layout engine via a combination of existing DOM interfaces and custom extensions in order to support visual state during time-travel (\S\ref{subsec:extensions}). 
Specifically, \sys creates checkpoints of the application's program and visual state (\S\ref{subsec:checkpoints}) at a configurable interval (2 seconds by default), and logs I/O and sources of nondeterminism (\S\ref{subsec:log}).
By regularly capturing checkpoints, \sys makes it possible to quickly time-travel to an arbitrary point in the
web application's execution.

\point{Debugging:}
When the developer encounters a bug, they can place and trigger a breakpoint to begin a debugging session.
At this point, the developer can use \sys to step forwards and backwards through the captured program execution to diagnose the bug.
To support stepping forwards, \sys uses its log to deterministically replay the program execution.

Stepping backwards is more involved (\S\ref{sec:debugger-features}), as \sys must return the application to a previous state.
To go back in time, \sys needs to return the application to a target JavaScript statement $s$ at a specific execution of the statement (at time $t$).
To track this information, \sys extends the JavaScript engine with the \emph{branch trace store} and \emph{timestamp store} performance monitors (\S\ref{subsec:perf-monitoring}).
\sys uses these performance monitors to determine $s$ and $t$.

\point{Time-travel:} To time-travel an application to statement $s$ at time $t$,
\sys loads the last checkpoint taken before $t$ and replays the log.
When execution is at the JavaScript event just prior to $t$, \sys enables the branch trace store and timestamp store, and places a conditional breakpoint on $s$ that triggers at time $t$.
Conditional breakpoints are a standard feature supported by all major JavaScript debuggers; the JavaScript engine will only trigger the breakpoint if $s$ executes at time $t$, which completes the time-travel operation.
%When there is no target time travel time, as is the case when the developer manually initializes replay from a particular snapshot, \sys must enable the performance monitors in order to respond to fine-grained time-travel requests at any given moment.

\point{Time-travel optimization:}
% OOPSLA REBUTTAL: The startup cost covers stepping within a single JavaScript event. Since a typical JavaScript event executes hundreds/thousands of JavaScript statements, it is common that a developer will step backwards hundreds of times without re-paying the startup cost.
% The startup cost is the amount of time it takes for McFly to place a new checkpoint just prior to the event that executes the statement that McFly is time-traveling to. Subsequent backward steps within that event do not pay the startup cost, as McFly steps backward by replaying execution using the new checkpoint.
% OOPSLA REBUTTAL: __C: Due to the small size of whole-application checkpoints, we did not need to implement this optimization in the prototype. We could implement the design described by Boothe to reduce the amount of checkpoints [6].
\sys opportunistically generates checkpoints \emph{during replay} to
reduce the latency of future time travel
operations. In particular, if \sys is time
traveling towards $t$, and must start from
a ``far-away'' checkpoint (where distance is
defined in terms of JavaScript events), \sys generates
a new checkpoint at the JavaScript event just prior to $t$. Thus, a sequence of stepping operations
within the same JavaScript event will use the new checkpoint; this is similar to an optimization by Boothe~\cite{boothe}.

\subsection{Supporting Visual State}
\label{subsec:extensions}

\sys supports visual state during debugging by checkpointing and logging changes to a high-level representation of the layout engine's visual state.
The layout engine already reveals much of its internal state in a high-level form to the JavaScript engine via the DOM.
However, some of the layout engine's internal state is not accessible via standard interfaces, including the state of animations on the web page and lists of active event listeners.
\sys requires access to this state to be able to checkpoint and deterministically re-execute applications.

\sys extends the layout engine with additional debugger-facing interfaces that provide read/write access to a high-level representation of internal layout engine state.
These extensions expose the same high-level state described in formal DOM specifications, which all web browsers adhere to~\cite{domSpec}.
For example, the DOM specification for HTTP request objects (\smtexttt{XMLHttpRequest}) describes network request objects as a state machine; \sys captures these network requests in terms of the internal state machine.
As a result, this architecture is \emph{portable} across all major web browsers.
\autoref{sec:implementation} details the specific extensions that our prototype version of \sys supports, as well as their implementation in a widely-used browser.

\subsection{Application Checkpoints}
\label{subsec:checkpoints}

\sys's web application checkpoints contain the application's program state (from the JavaScript engine) and visual state (from the layout engine).
As these two types of state are entangled through cross-references, \sys stores them together inside checkpoints as a single object graph.

\point{Program State:} At a high level, the program state within the JavaScript engine consists of the heap, the stack, and the program counter.
However, the JavaScript engine does not maintain a stack or a program counter between JavaScript events.
In addition, JavaScript events are typically short-lived (under a few milliseconds in duration) because JavaScript execution blocks UI interactions~\cite{jsmeter}; long-running events give the user the impression that the application is frozen.
This property is enforced by the web browser; if a JavaScript event runs for too long, the browser crashes the tab or raises an alert.

\sys \emph{defers} application checkpoints to occur between JavaScript events.
This design constrains where \sys can checkpoint the application, but does not unduly impact debugging performance since most events complete in under a millisecond.

This design ensures that the only program state that \sys needs to capture is the JavaScript heap, which all major web browsers can efficiently traverse using existing garbage collection routines.
When the traversal encounters an object in the heap that reflects and retains visual state in the layout engine, \sys serializes its high-level state.
To reinstate checkpointed program state, \sys uses internal interfaces present in all major JavaScript engines to reconstruct serialized objects.

\point{Visual State:} A web application's visual state consists of active GUI nodes on the webpage, inactive GUI nodes stored in the JavaScript heap, active CSS animations, browser-local persistent storage, the state of the random number generator, event listeners, the number of bytes consumed by each active HTML parser, and the state of network requests.
\sys serializes a high-level representation of these resources into the checkpoint's object graph using a combination of standard web interfaces and the extensions discussed in \autoref{subsec:extensions}.

\subsection{I/O and Nondeterminism Log}
\label{subsec:log}

\sys logs I/O and sources of nondeterminism that it uses to ensure that stepping operations are deterministic.
The log contains different types of entries, which each have a different logging and replay strategy.
We describe each type of log entry below; \autoref{sec:implementation} details which DOM interfaces use which strategy.

\point{Simple entries} correspond to synchronous interactions between JavaScript and the layout engine, such as querying for the date, that depend on browser-external state.
\sys logs these values during the original execution, and replays them while debugging.

\point{Event entries} correspond to JavaScript events.
\sys uses these to ensure that JavaScript events occur in the same order as the developer steps through the program.
During the original execution, \sys logs a high-level form of each event as it occurs.
At debug time, \sys replays events from the log.
\emph{This replay strategy reproduces event races observed during the original execution.}

\point{Inter-event visual state updates} occur between JavaScript events.
While JavaScript is executing, the layout engine defers certain visual state updates.
For example, layout engines only transition the internal state machine of HTTP request objects in quiescent periods between events.
During the original execution, \sys scans for and logs state changes for the relevant high-level state before every JavaScript event.
During replay, \sys applies the logged updates before the same event in order to keep the state in sync with JavaScript execution.

\point{Concurrent visual state updates} occur while JavaScript is executing.
The layout engine updates some visual state, such as CSS animations, concurrent with JavaScript execution.
Every time JavaScript code synchronously interacts with the layout engine, \sys scans for and logs any changes to concurrently updated state.
During replay, \sys prevents the layout engine from concurrently updating state and re-applies the state changes itself during the same synchronous interaction, which keeps the state in sync with JavaScript execution.

\sys uses a counter of synchronous interactions to denote a specific synchronous interaction in the log, and stores the counter value in checkpoints.
For example, a log entry with counter value 60 would be applied during the 60th synchronous layout engine interaction, which will be identical across deterministic replays.
This strategy preserves any data races between JavaScript code and the layout engine observed during the original execution.

%Note that these interactions execute in a critical section in the renderer, preventing the renderer from making state updates concurrent with the boundary crossing.
%Since \sys's log/replay actions occur within the same boundary crossing as the request it is interposing on, the renderer is unable to make state updates between the log/replay action and the original request, preventing races between \sys and the renderer.

\subsection{Debugger Features}
\label{sec:debugger-features}

\sys provides a full suite of
complements to existing debugger features, and exposes this functionality as an extension to a production JavaScript stepping debugger.
We only discuss \emph{step backwards} in this section; the remaining features are implemented similarly.

Step backwards complements \emph{step forwards}, and lets
the developer return to the previously-executed
program statement. 
Given that the debugger is paused at the statement $s$
at logical time $t=(c,b)$, where $c$ is the number of times the function has been called since enabling performance monitoring and $b$ is the number of backwards jumps (loop iterations) executed in the current function call,
the debugger must determine the statement and logical time of the previously-executed statement, $s^\prime$ and $t^\prime$:
\begin{itemize}[noitemsep,topsep=2pt]
  \item If $s$ is not the entry point of a basic block, then $s^\prime$ is the previous statement  in the block and $t^\prime = t$.
  \item If $s$ is the entry point of a basic block, then $s^\prime$ is the source statement of the \emph{previously taken branch}.
  \begin{itemize}[noitemsep,topsep=0pt]
    \item If $s^\prime$ is the current statement in the calling function, then $t^\prime$ is the logical time associated with the caller's call frame.
    \item Otherwise, $s^\prime$ is from the same function call as $s$. If $s^\prime$ is a loop header (e.g., \smtexttt{while(someCondition)}), then $s^\prime$ is from a previous loop iteration and $t^\prime=(c,b-1)$. Otherwise, $t^\prime = t$.
  \end{itemize} 
\end{itemize}
Finally, \sys places a conditional breakpoint
on $s^\prime$ that triggers when the logical time is $t^\prime$, and replays the program from the previous
checkpoint that is closest to the target logical time.
If the checkpoint is not close to the target time, \sys records a new checkpoint
just before the target JavaScript event in order to speed up subsequent step-back operations.

\subsection{Performance Monitors}
\label{subsec:perf-monitoring}
% http://www.intel.com/content/www/us/en/architecture-and-technology/64-ia-32-architectures-software-developer-system-programming-manual-325384.html
% https://msdn.microsoft.com/en-us/library/system.diagnostics.performancecounter(v=vs.110).aspx
% Can't find info on JVM perf counters other than a blog post, as they are "unsupported".
To replay execution to a specific statement at a particular
point in time, time-traveling debuggers need visibility into the current execution.
VM-based time-traveling debuggers typically use performance counters
on the processor for this purpose~\cite{intelSDM,clrPM}, but managed
languages, like JavaScript, lack comparable functionality.

\sys augments the JavaScript engine with two performance
monitors. The \textit{branch trace store} contains the
last branch instruction that was taken by each function that is
currently on the call stack.
The \textit{timestamp store}
contains the timestamp of each function on the call stack.
A timestamp is represented as the pair of the function's
call count since enabling performance monitoring and
the number of backwards jumps (loop iterations)
executed thus far in the function call.
For example, given the function \smtexttt{function a()\{while(true)\{\}\}},
the timestamp (3, 2) represents the second iteration of the loop
during the third call to the function.

\subsection{Replay Guarantees}

\sys guarantees that replay is identical to the original execution, including the data
returned from supported layout engine interfaces, the sequence of application-observed JavaScript events, and the pixels on the screen.
Animations may not move smoothly during replay, as the system fast-forwards animations to each observed state from the original execution whenever JavaScript code calls into the layout engine, but the JavaScript code will observe the same values seen in the original execution.
In addition, developers can use existing debugging tools, such as the DOM Explorer shown in \autoref{fig:visual-debug}, to inspect the GUI while debugging.

\begin{comment}
\subsection{Detecting Divergence}

\sys uses its log to detect replay divergence, which indicates that the application has encountered a source of nondeterminism or used a DOM interface that \sys does not handle.
\sys's log imposes a total order on I/O and nondeterminism events, which \sys uses to make various sanity checks during replay.
For example, \emph{simple entries} and \emph{inter-event visual state updates} always fall between \emph{event} entries.
If an event concludes before all of these entries replay, then replay has diverged.
Similarly, if a log entry does not make sense, e.g., the application registers a timer but the next log entry is for \smtexttt{Date}, then \sys knows that replay has diverged.
We used this feature to detect bugs in our prototype implementation of \sys.
\end{comment}

%% file: deterministic.tex
% !TeX root = main.tex

\begin{figure*}[tb]
\begin{tabularx}{\textwidth}{p{1.5cm} p{4.5cm} p{3cm} p{3.5cm} l}
\toprule
\normalsize\textbf{Resource} & \normalsize\textbf{Interface}  & \normalsize\textbf{High-level State} & \normalsize\textbf{Logged Data} & \normalsize\textbf{Type of Log Entries} \\ 
\midrule
Random & \texttt{Math.random()} & PRNG state & None & \\
Numbers & & & & \\
\midrule
Time & \texttt{Date} & Internal timestamps & Current time & Simple entries \\
\midrule
Timers & \texttt{setTimeout}, \texttt{setInterval} & Active timers & Timer IDs & Simple entries \\
\midrule
Events & \texttt{EventTarget} & Active event listeners & None &  \\
	& Indirect & None & Sequence of events & Event entries \\
\midrule
GUI & DOM & Live DOM nodes & Form changes, & Inter-event updates \\
    &     & & external resource loads, & Concurrent updates \\
    &     & & HTML parsing progress    & Concurrent updates \\
    & CSS Animations & Animation status & Animation advancement & Concurrent updates \\
\midrule
Network Requests & \texttt{XMLHttpRequest} & State machine & State machine changes & Inter-event updates \\
\midrule
Storage & \texttt{localStorage}, Cookies & Contents of storage & None &  \\
\bottomrule
\end{tabularx}

\caption{\textbf{The core browser interfaces that \sys supports.}
The table above summarizes the different resources that the browser's layout engine provides to JavaScript code, the high-level representation of their internal state, the data that \sys records into its log, and the types of log entries used.
In the ``interface'' column, \emph{Indirect} means that a program's
         interactions with the resource are implicit,
         i.e., the interactions do not use an explicit JavaScript interface.}
\label{fig:state-table}
\end{figure*}

\section{Implementation}
\label{sec:implementation}

We have implemented a prototype of \sys in Microsoft Edge.
This section describes in detail the changes we made to the layout engine to support \sys's checkpoints and logs (\S\ref{sec:js-proto}),
the changes to the JavaScript engine to support performance monitors (\S\ref{subsec:impl-perf-mon}),
%the changes to the debugger to support backwards stepping operations (\S\ref{subsec:debugger-feats}),
and the security implications of these changes (\S\ref{subsec:security-implications}).

\subsection{Layout Engine State}
\label{sec:js-proto}

Below, we walk through how our prototype implementation of \sys captures the high-level layout engine state described in \autoref{fig:state-table}.
Although the table lists many different resources, we only need to make a small number of modifications to the layout engine because standard browser interfaces already make a large amount of high-level state available to the debugger.

\point{Random numbers:} JavaScript applications use \smtexttt{Math.random()}
to generate random numbers, but cannot read or write
the internal state of the PRNG.
We modify the layout engine to let \sys query and reset the PRNG's state.
\sys stores the PRNG state in program checkpoints, and reinstates it
prior to replaying from a checkpoint.

\point{Current time:} The layout engine contains a \smtexttt{Date} interface that lets programs observe the current time as a \smtexttt{Date} object, which it queries from the OS.
We modify the layout engine to let \sys log and replay date requests.
To serialize \smtexttt{Date} objects into checkpoints, we use the existing \smtexttt{getTime()} function on the object to retrieve its timestamp.

\point{Timer status:} JavaScript applications create one-shot
timers via \smtexttt{setTimeout()}, and recurring timers via
\smtexttt{setInterval()}. Each timer is assigned a unique ID that
the layout engine arbitrarily determines. The layout engine does not
provide an interface for enumerating the set of active timers
or for controlling their IDs. We extend the layout engine to expose the set of active timers and to let \sys log and replay timer ID assignments.
We use the former modification to store active timers in checkpoints,
and the latter to deterministically replay timer IDs.
\sys deterministically replays the timer schedule as discussed under \emph{sequence of events}.

\point{Events:} JavaScript code can register
functions as handlers for events. For example, a program can register handlers for mouse click events.
Applications can register event listeners in three ways:
through properties on HTML tags (e.g.,
\smtexttt{<div onclick="a()">}), properties on
the DOM elements (e.g., \smtexttt{div.onclick=a;}),
or through the \smtexttt{addEventListener()} DOM interface (e.g.,
\smtexttt{div.addEventListener('click',a)}).
JavaScript code can enumerate event listeners
registered using the first two approaches, since the listeners
are reflected as properties on the associated DOM objects. However,
the layout engine does not let JavaScript code
enumerate handlers which were registered via \smtexttt{addEventListener()}.
Furthermore, the layout engine dispatches events to event handlers
in the order in which the handlers were registered, regardless
of the registration technique employed. The layout engine does not
expose this order, which must be recreated at replay time.

All DOM objects that generate events implement the
\smtexttt{EventTarget} interface.
We modify the layout engine to let \sys enumerate all event
handler information that is associated with an \smtexttt{EventTarget}.
\sys uses this extension to store handler
orders into checkpoints, and restore handler orders from checkpoints
using the preexisting handler registration interfaces.

\point{Sequence of events:} 
Each JavaScript execution context is single-threaded and completely event-driven, but the layout engine contains and controls
the JavaScript event queue.
The layout engine does not expose the queue to JavaScript code, but
events must be replayed in the same order as the original execution in order to
reproduce event races.
We extend the layout engine to let \sys intercept events
added to the event queue, which it uses to log and reproduce the original event order during a debugging session.

\point{GUI:} JavaScript code interacts with the GUI through the DOM tree.
Each HTML tag on a web page has a corresponding element object in the tree.
Each element object provides JavaScript code with read and write access to tag-specific state, such as the URL for
an \smtexttt{<img>} tag or the text in a form field.
We use existing JavaScript interfaces to serialize and deserialize the
entirety of the DOM tree into checkpoints.

\point{External resources:} A web page often includes external objects, e.g., HTML tags
which specify a \smtexttt{src} attribute and whose content
must be fetched from remote servers. The layout engine loads this content
in parallel with JavaScript execution on an I/O thread.
When the content finishes loading, the layout engine silently updates the applicable HTML element
with attributes, such as the \smtexttt{height} and \smtexttt{width}
of an image.
We modify the layout engine to let \sys log and replay network fetches.

\point{HTML parsing progress:} A web page contains one or more HTML documents, with secondary HTML documents appearing in frames.
The browser incrementally loads and parses HTML documents in parallel with JavaScript execution, which causes new nodes to appear in the DOM.
We extend the layout engine to expose the current byte offset in each document's parse stream.
\sys logs offset changes during the original execution.
During replay, \sys feeds the network stack the new bytes at the appropriate time (via the extension discussed in \emph{external resources}) and waits for the parser to consume them.

\point{CSS animations:} The DOM does not expose the CSS
animation state of an HTML tag---that state resides within
the layout engine. If \sys cannot read CSS animation state, then it cannot record an animation's progress
with respect to concurrently executing JavaScript code; this
would prevent \sys from faithfully recreating the
behavior. To enable high-fidelity replays of animations,
we modify the layout engine to let \sys read and write the frame counts corresponding to active CSS animations. Using this interface,
\sys can ``seek'' to a specific point in the
animation, and keep it synchronized with JavaScript execution.

\point{Connection status:} JavaScript applications communicate
with remote servers via \smtexttt{XMLHttpRequest} objects. Each
object encapsulates the state of a single HTTP request. At
logging time, the debugger can observe the state of each
request, including the content of the HTTP response, using existing methods on \smtexttt{XMLHttpRequest}
objects. The debugger needs a mechanism
to recreate logged \smtexttt{XMLHttpRequest} objects without creating
actual network connections. We extend the layout engine to let the debugger create \smtexttt{XMLHttpRequest} objects from
scratch, and set their internal state to arbitrary
values.

\point{Storage:} Web applications manage persistent local data
using cookies and the \smtexttt{localStorage} interface~\cite{domStorageSpec, domSpec}.
Both mechanisms export a key/value API. The layout engine
creates a separate storage area for each origin,
and prevents
different origins from accessing each other's data.
A page's origin is the 3-tuple of the protocol, hostname, and port
in the page's URL.
Since all of the active origins on a web page execute within the same JavaScript engine,
\sys can pose as any origin and manipulate its storage using
the same interfaces that are exposed to regular applications.

\point{Additional browser features:} \sys supports a core set of browser interfaces, which is sufficient to time-travel many existing web applications. 
Browsers regularly add new features, such as WebGL and Web Audio, that \sys does not support. 
\sys can be extended to support these features with additional browser modifications to expose their hidden state.

\subsection{Performance Monitors}
\label{subsec:impl-perf-mon}

For simplicity, we implement the \emph{branch trace store} and \emph{timestamp store} by
augmenting the browser's JavaScript interpreter.
When a performance monitor is enabled, we disable the
browser's JIT compiler, forcing JavaScript execution
to use the interpreter. As we mention in \autoref{subsec:tt-overview}, \sys only requires performance monitoring when a replayed
execution nears a target line of interest,
so this design has minimal performance impact.

\subsection{Security Implications}
\label{subsec:security-implications}

The layout engine modifications described in \S\ref{sec:js-proto} are only exposed to debugging tools.
They are not accessible to web applications via public JavaScript APIs.
These modifications do not affect the security model for web content---at logging time and during replay, browsers still use the same-origin policy to isolate content.

%% file: evaluation.tex
% !TeX root = main.tex
\section{Evaluation}
\label{sec:eval}

We evaluate \sys by running it on a corpus of web applications. Our evaluation addresses the following questions:
\begin{itemize}
	\item \textbf{Faithfulness}: Does \sys faithfully and deterministically re-execute web applications?
	\item \textbf{Performance}:	Does \sys support time-travel debugging at interactive speeds?
	\item \textbf{Overhead}: Does \sys impose acceptable overhead during normal web application execution?
\end{itemize}
We performed the evaluation on a desktop with a quad-core Intel Xeon E5-1620 clocked at 3.6 GHz with 16GB of RAM and a 7200 RPM SATA hard drive.

\subsection{Applications}

There are no established benchmark suites for time-traveling debuggers for web applications, so we collected one.
To perform a controlled evaluation, we chose applications that we had the source code to and that we could run locally in a non-production setting.
Conducting performance experiments on web applications running in production poses severe methodological challenges.
Every experiment is likely to capture a different version of the web application, due to updates or A/B testing, and with different content.
For example, Facebook regularly conducts experiments on its users that changes the code and presentation of the website, and Facebook's feed contains different advertisements, posts, and third-party code across visits.
%JWV: this is a risk, esp w/ the next sentence
%Production JavaScript code is also likely to be obfuscated (via minification), which mangles control flow and symbols and complicates the task of verifying that the debugger is working properly.
While we did not use production applications in this evaluation, we have verified that \sys works on production websites.

% i want you to drive home the point that running controlled experiments on live websites is hard
%you might want to mention obfuscated code, if that makes sense, since it clearly would have complicated verifying that the debugger was working properly. not sure if that’s a risk
%(of irritating someone)
%just image rob o’callahan reading this thing
%*imagine
% maybe say “While these benchmarks are not real, yada yada, we tested McFly on production websites (if true!) and verified that it works.”
%anything you can say to dispel the benchmark problem

We focus our qualitative and quantitative evaluation on benchmarks that exercise different components of \sys:

\begin{itemize}
	\item \bench{Delta-Blue}, \bench{Earley-Boyer}, \bench{RayTrace}, and \bench{Splay} are from the \emph{Octane} benchmark suite~\cite{octane}, and are memory intensive workloads that stress \sys's checkpoints. We modify the benchmarks to extend their runtime to $\sim$10 seconds to isolate \sys overhead from parsing/JIT warmup overhead. Unlike the other benchmarks, these programs have no I/O and are deterministic; we exclude these benchmarks from parts of the performance evaluation that use \sys's nondeterminism and I/O log.

	\item \bench{RayTrace (GUI)}~\cite{raytracegui} is the RayTrace program from the Octane benchmark suite with its original HTML GUI, which introduces I/O to the program.

	\item \bench{ColorGame}~\cite{colorgame} is an implementation of a test that demonstrates the \emph{Stroop effect}~\cite{stroop}. It uses AngularJS and jQuery, which are both complicated and commonly used libraries, and result in ColorGame having $\sim$3$\times$ as much code as the next largest benchmark. AngularJS exercises a wide variety of DOM features, and encodes crucial application data into the DOM directly. Thus, it is crucial that \sys correctly support this application's visual state during debugging sessions.

	\item \bench{CRUD} is a standard content management interface that uses jQuery to manage its user interface. CRUD uses HTML forms to let users create, update, and delete content, which \sys must correctly support for debugging to be deterministic.

	\item \bench{PacMan}~\cite{pacman} is an implementation of the classic Pac-Man game using the HTML5 canvas. It uses timers to update the contents of the canvas every $80$ms, and stresses \sys's ability to quickly serialize large DOM objects into checkpoints and log frequent events.
\end{itemize}
\autoref{tab:snapmetrics} describes the code size of each of these benchmarks, including HTML documents and JavaScript libraries. %Although ColorGame and CRUD are significantly larger than most of the other benchmarks due to their sizeable dependencies, they are representative of the complexity of existing web applications.

\subsection{Faithfulness}

We evaluated the faithfulness of \sys's time-travel debugging by using \sys to debug the benchmark applications.
While using breakpoints and \sys's stepping operations, we observed each application's visual and program states, and checked that it matched the original execution.

We manually verified that, across all of our benchmarks, \textbf{\sys faithfully and deterministically reproduces the program and visual states observed during the original execution.}
Using \sys, we were able to deterministically step forwards and backwards through web application executions while visual state updates, including those induced by CSS animations and network dependencies, remained synchronized with JavaScript execution.

\begin{figure*}[t]
\centering
\begin{tabular}{l r r r r r r r r}
\toprule
& & \multicolumn{1}{c}{\textbf{Process}} & \multicolumn{1}{c}{\textbf{\# JS}} & \multicolumn{5}{c}{\textbf{Checkpoint}} \\
\multicolumn{1}{c}{\textbf{Program}}              & \multicolumn{1}{c}{\textbf{Code}}               & \multicolumn{1}{c}{\textbf{Size}}      & \multicolumn{1}{c}{\textbf{Objs}}         & \multicolumn{1}{c}{\textbf{Create}}         & \multicolumn{1}{c}{\textbf{Write}}        & \multicolumn{1}{c}{\textbf{Resume}}      & \multicolumn{1}{c}{\textbf{Size}}        & \multicolumn{1}{c}{\textbf{Compressed}} \\
\midrule
\bench{Color-Game}    & $746$\niceunitkb   & $44$\niceunitmb   & $18$\niceunitk  & $0.12$\niceunits  & $0.12$\niceunits  & $0.43$\niceunits  & $3.3$\niceunitmb   & $0.9$\niceunitmb \\
\bench{CRUD}          & $250$\niceunitkb   & $28$\niceunitmb   & $14$\niceunitk  & $0.11$\niceunits  & $0.10$\niceunits  & $0.37$\niceunits  & $2.4$\niceunitmb   & $0.5$\niceunitmb\\
\bench{Delta-Blue}    & $36$\niceunitkb    & $34$\niceunitmb   & $13$\niceunitk  & $0.09$\niceunits  & $0.17$\niceunits  & $0.36$\niceunits  & $2.1$\niceunitmb   & $0.4$\niceunitmb\\
\bench{Earley-Boyer}  & $244$\niceunitkb   & $123$\niceunitmb  & $17$\niceunitk  & $0.09$\niceunits  & $0.11$\niceunits  & $0.30$\niceunits  & $3.3$\niceunitmb   & $0.6$\niceunitmb\\
\bench{PacMan}        & $50$\niceunitkb    & $31$\niceunitmb   & $13$\niceunitk  & $0.13$\niceunits  & $0.14$\niceunits  & $0.45$\niceunits  & $2.2$\niceunitmb   & $0.5$\niceunitmb\\
\bench{RayTrace}      & $38$\niceunitkb    & $73$\niceunitmb   & $12$\niceunitk  & $0.06$\niceunits  & $0.09$\niceunits  & $0.28$\niceunits  & $2.1$\niceunitmb   & $0.4$\niceunitmb\\
\bench{Splay}         & $17$\niceunitkb    & $538$\niceunitmb  & $12$\niceunitk  & $0.08$\niceunits  & $0.10$\niceunits  & $0.33$\niceunits  & $2.3$\niceunitmb   & $0.5$\niceunitmb\\
\midrule
\bench{Average} & $197$\niceunitkb & $124$\niceunitmb & $14$\niceunitk & $0.10$\niceunits & $0.12$\niceunits & $0.36$\niceunits & $2.5$\niceunitmb & $0.5$\niceunitmb \\
\bottomrule
\end{tabular}
\caption{\textbf{\sys takes a fraction of a second to create or restore a checkpoint.} Checkpoints are also significantly smaller than the size of the browser process they capture.}
\label{tab:snapmetrics}
\end{figure*}

\begin{figure}[t]
\centering
\begin{tabularx}{0.43\textwidth}{l r r}
\toprule
\textbf{Program}             & \textbf{Overhead}          & \textbf{Startup} \\
\midrule
\bench{Color-Game}    & $4$\niceunitpct   & $4.5$\niceunits  \\
\bench{CRUD}          & $0$\niceunitpct         & $3.2$\niceunits   \\
\bench{PacMan}        & $6$\niceunitpct   & $4.7$\niceunits   \\
\bench{RayTrace (GUI)}   & $5$\niceunitpct   & $2.9$\niceunits   \\
\midrule
\bench{Average} & $4$\niceunitpct & $3.8$\niceunits \\
\bottomrule
\end{tabularx}
\caption{\textbf{Reverse-debugging overhead}: Overall, \sys imposes low overhead on our benchmark applications (up to 6\% w/ checkpoints every 2 seconds) and requires a short one-time startup cost to initialize efficient reverse-step debugging.}
\label{tab:ttdmetrics}
\end{figure}

\subsection{Performance}
\label{sec:snapshot-perf}

In order to be useful, \sys must step through an execution of a web at interactive speeds.
While stepping forwards is straightforward and involves deterministic replay, stepping backwards in time involves more costly operations.
Specifically, stepping backwards involves resuming execution from the nearest checkpoint and playing forwards to the JavaScript statement of interest.
In addition, the first time the developer steps backwards, the debugger creates a checkpoint just prior to the JavaScript event that executes the statement of interest.

\sys's step backwards overhead has two components:
\begin{itemize}
	\item \textbf{Startup Cost}: The first time the developer steps backwards, \sys replays the execution from the nearest checkpoint and creates a new checkpoint just prior to the JavaScript event of interest. 
	\item \textbf{Resuming Execution from Checkpoint}: Once the startup cost is paid, the cost of subsequent step backwards operations within the same JavaScript event is dominated by the time to resume from the newly created checkpoint.
\end{itemize}

We drive each benchmark through a fixed series of events using a PowerShell script that provides application inputs. Each script lasts approximately 10 seconds.
We run all benchmark programs and checkpoint their state \emph{every second} for the checkpoint cost evaluation in order to collect more data points, and \emph{every two seconds} for the startup cost evaluation to reflect a more representative value for everyday usage.
We calculate the average time to resume from a checkpoint (from disk), and the time to take the first backwards step from 10 random breakpoints.
From our results, we observe the following:

\point{\sys's stepping operations run at interactive speeds in the common case.}
After paying a one-time startup fee, the time to execute a backwards step in \sys is dominated by the time it takes to resume from the nearest checkpoint (0.36s on average, as shown in \autoref{tab:snapmetrics}).

\point{\sys imposes an acceptable backwards step startup cost.}
This startup cost is, on average, 3.8 seconds on our benchmark applications or roughly twice the checkpoint rate (\autoref{tab:ttdmetrics}).

%\begin{itemize}[topsep=0pt,noitemsep,leftmargin=10pt]
%  \item \textbf{Process Size:} Virtual memory used by browser process
%  \item \textbf{JS Objs:} \# of JavaScript objects in checkpoint
%  \item \textbf{Create:} Time to produce the checkpoint
%  \item \textbf{Write:} Time to write checkpoint to disk
%  \item \textbf{Resume:} Time to resume from checkpoint on disk
%  \item \textbf{Checkpoint Size:} Size of checkpoint in memory
%  \item \textbf{File Size:} Size of compressed checkpoint on disk
%\end{itemize}
%\autoref{tab:snapmetrics} displays the results of these experiments. From our results, we conclude the following:

\subsection{Overhead}

To be usable, \sys must not impose significant time and space overheads during normal web application execution.
The following metrics contribute to \sys's runtime and space overheads:
\begin{itemize}
	\item \textbf{Log Growth}: The growth rate of the nondeterminism and I/O log. Since a fast-growing log will exhaust disk and memory resources, this metric bounds the practical duration of program executions that \sys can support.
	\item \textbf{Checkpoint Size}: The size of application checkpoints, which \sys takes at a regular and configurable interval during the original execution. If checkpoints are large, then it will be impractical to take frequent checkpoints, which increases the initial cost return to a specific point in an execution. The \emph{compressed} size indicates the checkpoint's size on disk when compressed with gzip.
	\item \textbf{Checkpoint Creation Time}: The amount of time it takes to create a checkpoint. If it takes a long time to create a checkpoint, then \sys will induce noticeable slowdowns during the initial execution of the program.
\end{itemize}

To measure these, we again drive each benchmark through a fixed series of events using a PowerShell script for approximately 10 seconds.
For checkpoint operations, we run each benchmark in a configuration that takes a checkpoint every second.
For log growth, we run each benchmark without taking any checkpoints, maximizing log size.
For overall overhead during a normal execution, we measure the runtime of each benchmark in a configuration that takes a checkpoint every two seconds, the default configuration, and compare with the benchmark's runtime without \sys.

\noindent As \autoref{tab:snapmetrics} shows, \textbf{checkpoint creation takes less than an eighth of a second} on our benchmark applications. \sys takes an average of 100 milliseconds to create a checkpoint and 120 milliseconds to serialize the checkpoint to disk.
\sys's checkpoint operations are fast enough to support frequent checkpoints with acceptable overhead.
%Note that these operations could be further optimized by tightly integrating with Chakra's generational garbage collector~\cite{tardis}.
% These operations could be further optimized by piggybacking on generational garbage collector operations to avoid crawling and checkpointing the entire JavaScript heap~\cite{tardis}.

\point{\sys checkpoints compress to less than a megabyte} on our benchmark applications (\autoref{tab:snapmetrics}).
\sys checkpoints contain the web application's complete state as a lightweight high-level representation that is amenable to compression.
Compressed \sys checkpoints are two orders of magnitude smaller than the browser's memory footprint at the process-level.

\begin{figure}[t]
\centering
\begin{tabular}{l r c}
\toprule
\textbf{Program}               & \textbf{Log Growth} \\
\midrule
\bench{Color-Game}    & $0.6$\niceunitkbs \\
\bench{CRUD}          & $0.2$\niceunitkbs \\
\bench{PacMan}        & $1.5$\niceunitkbs \\
\bench{RayTrace (GUI)}   & $0.9$\niceunitkbs \\
\midrule
\bench{Average} & $0.8$\niceunitkbs \\
\bottomrule
\end{tabular}
\caption{\textbf{Log overhead}: \sys's \emph{uncompressed} nondeterminism and I/O log grows slowly.}
\label{tab:logmetrics}
\end{figure}

\point{\sys's nondeterminism and I/O log grows at less than 2KB/s} (\autoref{tab:logmetrics}).
The benchmark with the most nondeterminism, PacMan, has the largest log growth rate of 1.5KB/s.
At that rate, \sys could record PacMan's execution for \emph{over 11 years} on a 500GB hard drive.
%Reported log growth rates are \emph{uncompressed}; compression would yield even smaller logs.

\point{With checkpoints every two seconds, \sys's overall overhead is 4\% on average} on our benchmark applications (\autoref{tab:ttdmetrics}).
In some cases, such as CRUD, \sys imposes \emph{no} overhead because checkpoints occur between JavaScript events, when the application is waiting for user input.
\sys can impose even lower runtime overheads in exchange for longer step backwards startup costs with a lower checkpoint rate during execution.

%% file: related-work.tex
% !TeX root = main.tex

% TODO: Add concurrency work???
% RR: https://github.com/mozilla/rr/blob/5dc10d6e11e5b648913388873d1c66c0f009f29b/src/ReplayTimeline.cc#L1405
\begin{figure*}[tb]
\centering\begin{tabular}{l c c c c}
	\toprule
	 & \textbf{Program} & & \textbf{Supports Debugging} & \textbf{Step-backward} \\
	\textbf{System} & \textbf{Replay} & \textbf{GUI Support} & \textbf{Web Applications} & \textbf{Support} \\
	\midrule	
	\textbf{Record and replay systems:} \\
	Timelapse~\cite{timelapse} & \cmark & $\text{\cmark}^{*}$ & \cmark  \\
	Mugshot~\cite{mugshot} & \cmark & $\text{\cmark}^{*}$ & \cmark \\
	Jalangi~\cite{jalangi} & \cmark & & \cmark &\\
	% Cost: O(L) or w/e for RR - dominated by log size (or trace size / runtime?)
	\midrule
	\textbf{Time-traveling debuggers:} \\
	RR~\cite{rrtool} & \cmark & &  & \cmark \\
	Jardis~\cite{jardis} & \cmark & & & \cmark \\
	\midrule
	% Cost: O(H) for sys - dominated by heap cost
	\sys & \cmark & \cmark & \cmark & \cmark \\
	\bottomrule
\end{tabular}
	\caption{\textbf{\sys is the first time-traveling debugger for web applications.} No prior debugger is able to support step-backward debugger commands with GUI support at interactive speed. * indicates that the system does not correctly reproduce data races between the layout engine and JavaScript.}
	\label{fig:prior-work}
\end{figure*}

\section{Related Work}
\label{sec:relwork}

Although \sys is the first time-traveling debugger for web applications, time-traveling debuggers exist in several other non-graphical settings (\S\ref{subsec:rl:ttd}).
Plain record-and-replay systems exist for GUI applications, but these are  unable to support time-travel debugging at interactive speeds because they cannot checkpoint and roll back program and visual state (\S\ref{subsec:rl:dr}).
\autoref{fig:prior-work} summarizes prior work that supports replaying web application executions.

\subsection{Time-Travel Debugging}
\label{subsec:rl:ttd}

\sys is the first time-traveling debugger for web applications.
Previous time-traveling debuggers for other settings fall into three main categories: \emph{application-level} debuggers, \emph{VM-level} debuggers, and \emph{omniscient} debuggers.

\point{Application-level:}
Tardis~\cite{tardis}, Jardis~\cite{jardis}, UndoDB~\cite{undodb}, Boothe~\cite{boothe}, and RR~\cite{rrtool} record and replay program interactions with a well-defined interface to an external environment, but do not recreate state in the external environment during debugging.
In other words, these debuggers do not recreate a GUI application's visual state.
%In other words, these debuggers interpose on boundaries that are equivalent in fidelity to the script boundary described in \autoref{sec:intro}, so state beyond the boundary, such as the program's GUI, is not visible or debuggable during replay.
Tardis and Jardis debug .NET CLR and Node.js programs respectively, and replays interactions with native (C/C++) methods.
UndoDB, RR, and Boothe debug the user space of processes, and replay interactions with the kernel and hardware.
Boothe and RR use a similar optimization as \sys to recreate checkpoints during replay to amortize time-travel cost. 
RR can step forwards and backwards through a Firefox execution, but it is designed to debug the browser itself rather than web applications and imposes single-threaded browser execution at all times.
%is one to two orders of magnitude slower than \sys during program execution because it imposes single-threaded browser execution~\cite{rr-perf}.
%, logs and replays 
%but is intended to debug the browser itself, not web applications that run atop a browser, so rr logs and replays comparatively large amount of state, and is much more complex than ReJS.
%this capability cannot be used to debug web applications running inside of Firefox 
%While external state is not important to debugging many native applications, web developers expect to be able to inspect DOM elements during replay using existing development tools (\autoref{fig:breakpoint-debug}); \sys provides developers with this functionality.

\point{VM-level:}
VM-level hypervisors like XenTT~\cite{burtsev}, ReVirt~\cite{revirt,dunlap-revirt-2}, ReTrace~\cite{retrace}, and TTVM~\cite{kingVM} can time-travel entire virtual machines, but at the expense of large program traces and slow time-travel.
Reverse-step debugging, like that provided by \sys, requires time-travel at interactive speeds to be practical.

\point{Omniscient:} Omniscient debuggers provide time-traveling features by
recording program state changes after every instruction, which produces
large program traces and imposes high overhead during execution.
Examples of omniscient debuggers include Chronon~\cite{chronon}, TOD~\cite{omniscientTOD}, ODB~\cite{odb}, and Tralfamadore~\cite{tralfamadore}.

\subsection{Deterministic Replay}
\label{subsec:rl:dr}

Pure deterministic replay systems can record and replay an application's execution, but do not support periodic checkpoints or reverse debugging.
As a result, these systems are unable to support backwards stepping operations at interactive speeds.
We center our discussion on three different runtime environments.

\point{Browser:} % https://trac.webkit.org/wiki/WebReplayMechanics bugs for TL.
Mugshot~\cite{mugshot} and Timelapse~\cite{timelapse} deterministically replay web application executions by recording and replaying the event schedule and I/O operations; we compare these systems to \sys in \autoref{fig:prior-work}.
%between the JavaScript engine and the browser to record and replay web applications.
To accomplish this goal, Mugshot uses program rewriting and JavaScript reflection while Timelapse modifies the WebKit layout engine.
While \sys also modifies the layout engine, Timelapse's ``hypervisor-like record/replay strategy relies on the layered architecture of WebKit'' and is not portable to other browsers; in contrast, \sys's architecture builds on web standards that are supported across all major browsers.
Neither system is able to provide step-backward debugger commands at interactive speeds because they are unable to capture application checkpoints.
Furthermore, Mugshot and Timelapse do not support layout engine operations that mutate visual state in parallel with JavaScript execution, such as CSS animations, which can cause divergent application replays.

Jalangi~\cite{jalangi} supports selectively recording and replaying a subset of a program's code in support of dynamic analyses.
On the user-selected subset of code, Jalangi logs and replays interactions with the browser's native functions with considerable overhead (26X during recording and 30x during replay), and does not support visual state during replay.

%assume that replaying the same JavaScript events and network responses will result in the same visual state during replay, 
% and do not support data races between JavaScript execution and the layout engine.
%Thus, they are unable to accurately replay applications that exhibit the data races discussed in \S\ref{subsec:js-races}.

\point{Android:}
The Android runtime environment is similar to the browser environment in that applications are event-driven and use a single thread to update the GUI.
Valera~\cite{valera} and ReRan~\cite{reran} interpose on the interface between Android applications and the Android platform to capture nondeterministic event schedules and I/O operations.
%Like record-and-replay systems for web applications, these systems assume that interactions with Android's GUI interface are deterministic, and do not handle the possibility of data races with the environment.

\point{JVM:}
JVM applications communicate with the environment and internal JVM components via native methods.
Existing record-and-replay systems for the JVM treat state below the native methods, such as visual state, as external to replay.
DejaVu assumes all native methods are deterministic, preventing applications from using nondeterministic APIs~\cite{dejavu}.
ORDER records and replays select nondeterministic APIs, preventing developers from inspecting or observing JVM-external state, like the GUI, during replay~\cite{order}.

%\point{Process-level}:
%Deterministic replay systems for arbitrary processes.

% JaRec: ???
%[13] A. Georges, M. Christiaens, M. Ronsse, and K. De Bosschere.
%Jarec: a portable record/replay environment for multi-threaded
%java applications. Softw. Pract. Exper., 34(6):523–547, 2004.

% Eidetic systems:
% - No whole-program checkpoints, just checkpoints at process creation.
% - No support for recreating state in external environment.
% Chimera: Hybrid Program Analysis for Determinism, PLDI 12, Peter chen.
% - No whole-program checkpoints, just checkpoints at process creation.
% - No support for recreating state in external environment.
% - Requires browser re-compiled to add weak locks around racy code.
% DoublePlay: parallelizing sequential logging and replay, ASPLOS 2011, Peter Chen.
% - Does not reproduce external state (OS)
% Respec: efficient online multiprocessor replay via speculation and external determinism, ASPLOS 2010, Peter Chen.
% - Doesn't recreate external environment.
% PRES: probabilistic replay with execution sketching on multiprocessors. SOSP 2009, Yuanyuan Zhou.
% - Replays may not be identical, which is needed for TTD.
% - As far as I can tell... doesn't recreate external state...????
% Flashback: A Light-weight Extension for Rollback and Deterministic Replay for Software Debugging. The  Proceedings of the annual Usenix technical conference (USENIX'04), Yuanyuan Zhou
% - Captures and replays system calls.

% iReplayer supports in-situ replay for a limited epoch; some sys calls are delayed to let syscalls deterministically re-execute during epoch.

%% file: conclusion.tex
% !TeX root = main.tex
\section{Conclusion}

This paper presents \sys, the first time-traveling debugger for web applications.
\sys provides accurate time-travel debugging that maintains JavaScript and visual state in sync at all times.
We show that \sys lets developers freely step forwards \emph{and backwards} through a web application's execution at interactive speed.
%\sys imposes low overheads during execution . %\sys produces small logs and program checkpoints.
Core parts of \sys have been incorporated into a time-traveling debugger product from Microsoft~\cite{chakra-ttd}.

%This paper presents \bleak, the first effective system for debugging
%client-side memory leaks in web applications. We show that \bleak has
%high precision and finds numerous previously-unknown memory leaks in
%web applications and libraries. BLeak is open source~\cite{bleak-github}, and is available for download at \url{http://bleak-detector.org/}.

%We believe the insights we develop for \bleak are applicable to a broad class of GUI applications, including mobile applications.
%Many mobile applications are actually hybrid applications, which combine both native and browser components.
%However, even native GUI applications, like web applications, are commonly event-driven and repeatedly visit specific views.
%We plan in future work to explore the application of \bleak's techniques to find memory leaks in GUI applications.

%% file: acknowledgements.tex
\section*{Acknowledgements}

We would like to thank Nikhil Khandelwal and Rob Paveza for
their initial drive to start this project in Chakra.
Akrosh Gandhi, Sandeep Agarwal, Arunesh Chandra, and Fiona Fung provided extensive help working in the Chakra codebase and managing the project.
Edgardo Zoppi contributed during an internship at Microsoft Research.
Priyank Desai, James Lissiak, and Rob Lourens helped add prototype time-travel support to the F12 and Visual Studio Code debuggers.
Finally, we wish to thank Ed Maurer, Gaurav Seth, Dan Moseley, and Andy Sterland, who supported the development of the JavaScript component of McFly that is used to time-travel Node.js applications.

% who have continued developing the debugger for Node.js applications.
%On the acknowledgements I think we should take from the list of Acks from the Jardis paper + authors and Andy Sterland:
%Ed Maurer
%Gaurav Seth
%Dan Moseley
%Andy Sterland

%% file: main.bbl
%%% -*-BibTeX-*-
%%% Do NOT edit. File created by BibTeX with style
%%% ACM-Reference-Format-Journals [18-Jan-2012].

\begin{thebibliography}{46}

%%% ====================================================================
%%% NOTE TO THE USER: you can override these defaults by providing
%%% customized versions of any of these macros before the \bibliography
%%% command.  Each of them MUST provide its own final punctuation,
%%% except for \shownote{}, \showDOI{}, and \showURL{}.  The latter two
%%% do not use final punctuation, in order to avoid confusing it with
%%% the Web address.
%%%
%%% To suppress output of a particular field, define its macro to expand
%%% to an empty string, or better, \unskip, like this:
%%%
%%% \newcommand{\showDOI}[1]{\unskip}   % LaTeX syntax
%%%
%%% \def \showDOI #1{\unskip}           % plain TeX syntax
%%%
%%% ====================================================================

\ifx \showCODEN    \undefined \def \showCODEN     #1{\unskip}     \fi
\ifx \showDOI      \undefined \def \showDOI       #1{#1}\fi
\ifx \showISBNx    \undefined \def \showISBNx     #1{\unskip}     \fi
\ifx \showISBNxiii \undefined \def \showISBNxiii  #1{\unskip}     \fi
\ifx \showISSN     \undefined \def \showISSN      #1{\unskip}     \fi
\ifx \showLCCN     \undefined \def \showLCCN      #1{\unskip}     \fi
\ifx \shownote     \undefined \def \shownote      #1{#1}          \fi
\ifx \showarticletitle \undefined \def \showarticletitle #1{#1}   \fi
\ifx \showURL      \undefined \def \showURL       {\relax}        \fi
% The following commands are used for tagged output and should be
% invisible to TeX
\providecommand\bibfield[2]{#2}
\providecommand\bibinfo[2]{#2}
\providecommand\natexlab[1]{#1}
\providecommand\showeprint[2][]{arXiv:#2}

\bibitem[\protect\citeauthoryear{{Apple Inc.}}{{Apple Inc.}}{2018a}]%
        {safari-debugger}
\bibfield{author}{\bibinfo{person}{{Apple Inc.}}}
  \bibinfo{year}{2018}\natexlab{a}.
\newblock \bibinfo{title}{Tools - Safari for Developers}.
\newblock
  \bibinfo{howpublished}{\url{https://developer.apple.com/safari/tools/}}.
\newblock
\newblock
\shownote{[Online; accessed 10-April-2018].}


\bibitem[\protect\citeauthoryear{{Apple Inc.}}{{Apple Inc.}}{2018b}]%
        {webkit}
\bibfield{author}{\bibinfo{person}{{Apple Inc.}}}
  \bibinfo{year}{2018}\natexlab{b}.
\newblock \bibinfo{title}{WebKit}.
\newblock \bibinfo{howpublished}{\url{https://webkit.org/}}.
\newblock
\newblock
\shownote{[Online; accessed 10-April-2018].}


\bibitem[\protect\citeauthoryear{Barr and Marron}{Barr and Marron}{2014}]%
        {tardis}
\bibfield{author}{\bibinfo{person}{Earl~T. Barr} {and} \bibinfo{person}{Mark
  Marron}.} \bibinfo{year}{2014}\natexlab{}.
\newblock \showarticletitle{{Tardis: Affordable time-travel debugging in
  managed runtimes}}. In \bibinfo{booktitle}{\emph{Proceedings of the 2014
  Conference on Object-Oriented Programming, Systems, Languages, and
  Applications}}. \bibinfo{pages}{67--82}.
\newblock
\urldef\tempurl%
\url{https://doi.org/10.1145/2660193.2660209}
\showDOI{\tempurl}


\bibitem[\protect\citeauthoryear{Barr, Marron, Maurer, Moseley, and Seth}{Barr
  et~al\mbox{.}}{2016}]%
        {jardis}
\bibfield{author}{\bibinfo{person}{Earl~T. Barr}, \bibinfo{person}{Mark
  Marron}, \bibinfo{person}{Ed Maurer}, \bibinfo{person}{Dan Moseley}, {and}
  \bibinfo{person}{Gaurav Seth}.} \bibinfo{year}{2016}\natexlab{}.
\newblock \showarticletitle{Time-travel debugging for JavaScript/Node.js}. In
  \bibinfo{booktitle}{\emph{Proceedings of the 24th {ACM} {SIGSOFT}
  International Symposium on Foundations of Software Engineering}}.
  \bibinfo{pages}{1003--1007}.
\newblock
\urldef\tempurl%
\url{https://doi.org/10.1145/2950290.2983933}
\showDOI{\tempurl}


\bibitem[\protect\citeauthoryear{Basques}{Basques}{2018}]%
        {chrome-debugger}
\bibfield{author}{\bibinfo{person}{Kayce Basques}.}
  \bibinfo{year}{2018}\natexlab{}.
\newblock \bibinfo{title}{{Get Started with Debugging JavaScript in Chrome
  DevTools}}.
\newblock
  \bibinfo{howpublished}{\url{https://developers.google.com/web/tools/chrome-devtools/javascript/}}.
\newblock
\newblock
\shownote{[Online; accessed 10-April-2018].}


\bibitem[\protect\citeauthoryear{Beller, Spruit, Spinellis, and Zaidman}{Beller
  et~al\mbox{.}}{2018}]%
        {debugger-quotes}
\bibfield{author}{\bibinfo{person}{Moritz Beller}, \bibinfo{person}{Niels
  Spruit}, \bibinfo{person}{Diomidis Spinellis}, {and} \bibinfo{person}{Andy
  Zaidman}.} \bibinfo{year}{2018}\natexlab{}.
\newblock \showarticletitle{{On the Dichotomy of Debugging Behavior Among
  Programmers}}. In \bibinfo{booktitle}{\emph{Proceedings of the 40th
  International Conference on Software Engineering}}.
\newblock
\urldef\tempurl%
\url{https://doi.org/10.1145/3180155.3180175}
\showDOI{\tempurl}


\bibitem[\protect\citeauthoryear{Boothe}{Boothe}{2000}]%
        {boothe}
\bibfield{author}{\bibinfo{person}{Bob Boothe}.}
  \bibinfo{year}{2000}\natexlab{}.
\newblock \showarticletitle{Efficient algorithms for bidirectional debugging}.
  In \bibinfo{booktitle}{\emph{Proceedings of the 2000 {ACM} {SIGPLAN}
  Conference on Programming Language Design and Implementation}}.
  \bibinfo{pages}{299--310}.
\newblock
\urldef\tempurl%
\url{https://doi.org/10.1145/349299.349339}
\showDOI{\tempurl}


\bibitem[\protect\citeauthoryear{Burg, Bailey, Ko, and Ernst}{Burg
  et~al\mbox{.}}{2013}]%
        {timelapse}
\bibfield{author}{\bibinfo{person}{Brian Burg}, \bibinfo{person}{Richard
  Bailey}, \bibinfo{person}{Andrew~J. Ko}, {and} \bibinfo{person}{Michael~D.
  Ernst}.} \bibinfo{year}{2013}\natexlab{}.
\newblock \showarticletitle{Interactive record/replay for web application
  debugging}. In \bibinfo{booktitle}{\emph{Proceedings of the 26th Symposium on
  User Interface Software and Technology}}. \bibinfo{pages}{473--484}.
\newblock
\urldef\tempurl%
\url{https://doi.org/10.1145/2501988.2502050}
\showDOI{\tempurl}


\bibitem[\protect\citeauthoryear{Burmister}{Burmister}{2012}]%
        {raytracegui}
\bibfield{author}{\bibinfo{person}{Adam Burmister}.}
  \bibinfo{year}{2012}\natexlab{}.
\newblock \bibinfo{title}{{Flog.RayTracer Canvas Demo}}.
\newblock
  \bibinfo{howpublished}{\url{https://web.archive.org/web/20120218103726/http://labs.flog.co.nz:80/raytracer}}.
\newblock
\newblock
\shownote{[Online; accessed 16-April-2018].}


\bibitem[\protect\citeauthoryear{Burtsev, Johnson, Hibler, Eide, and
  Regehr}{Burtsev et~al\mbox{.}}{2016}]%
        {burtsev}
\bibfield{author}{\bibinfo{person}{Anton Burtsev}, \bibinfo{person}{David
  Johnson}, \bibinfo{person}{Mike Hibler}, \bibinfo{person}{Eric Eide}, {and}
  \bibinfo{person}{John Regehr}.} \bibinfo{year}{2016}\natexlab{}.
\newblock \showarticletitle{{Abstractions for Practical Virtual Machine
  Replay}}. In \bibinfo{booktitle}{\emph{Proceedings of the 12th Conference on
  Virtual Execution Environments}}. \bibinfo{pages}{93--106}.
\newblock
\urldef\tempurl%
\url{https://doi.org/10.1145/2892242.2892257}
\showDOI{\tempurl}


\bibitem[\protect\citeauthoryear{Choi and Srinivasan}{Choi and
  Srinivasan}{1998}]%
        {dejavu}
\bibfield{author}{\bibinfo{person}{Jong-Deok Choi} {and}
  \bibinfo{person}{Harini Srinivasan}.} \bibinfo{year}{1998}\natexlab{}.
\newblock \showarticletitle{{Deterministic Replay of Java Multithreaded
  Applications}}. In \bibinfo{booktitle}{\emph{Proceedings of the SIGMETRICS
  Symposium on Parallel and Distributed Tools}}. \bibinfo{publisher}{ACM},
  \bibinfo{address}{New York, NY, USA}, \bibinfo{pages}{48--59}.
\newblock
\showISBNx{1-58113-001-5}
\urldef\tempurl%
\url{https://doi.org/10.1145/281035.281041}
\showDOI{\tempurl}


\bibitem[\protect\citeauthoryear{{Chromium Contributors}}{{Chromium
  Contributors}}{2018}]%
        {blink}
\bibfield{author}{\bibinfo{person}{{Chromium Contributors}}.}
  \bibinfo{year}{2018}\natexlab{}.
\newblock \bibinfo{title}{{Blink - The Chromium Projects}}.
\newblock \bibinfo{howpublished}{\url{https://www.chromium.org/blink}}.
\newblock
\newblock
\shownote{[Online; accessed 10-April-2018].}


\bibitem[\protect\citeauthoryear{{Chronon Systems}}{{Chronon Systems}}{2017}]%
        {chronon}
\bibfield{author}{\bibinfo{person}{{Chronon Systems}}.}
  \bibinfo{year}{2017}\natexlab{}.
\newblock \bibinfo{title}{{Chronon, a DVR for Java}}.
\newblock \bibinfo{howpublished}{\url{http://chrononsystems.com}}.
\newblock
\newblock
\shownote{[Online; accessed 30-April-2017].}


\bibitem[\protect\citeauthoryear{Dunlap, King, Cinar, Basrai, and Chen}{Dunlap
  et~al\mbox{.}}{2002}]%
        {revirt}
\bibfield{author}{\bibinfo{person}{George~W. Dunlap},
  \bibinfo{person}{Samuel~T. King}, \bibinfo{person}{Sukru Cinar},
  \bibinfo{person}{Murtaza~A. Basrai}, {and} \bibinfo{person}{Peter~M. Chen}.}
  \bibinfo{year}{2002}\natexlab{}.
\newblock \showarticletitle{{ReVirt: Enabling Intrusion Analysis Through
  Virtual-Machine Logging and Replay}}. In
  \bibinfo{booktitle}{\emph{Proceedings of the 5th Symposium on Operating
  System Design and Implementation}}.
\newblock
\urldef\tempurl%
\url{http://www.usenix.org/events/osdi02/tech/dunlap.html}
\showURL{%
\tempurl}


\bibitem[\protect\citeauthoryear{Dunlap, Lucchetti, Fetterman, and Chen}{Dunlap
  et~al\mbox{.}}{2008}]%
        {dunlap-revirt-2}
\bibfield{author}{\bibinfo{person}{George~W. Dunlap},
  \bibinfo{person}{Dominic~G. Lucchetti}, \bibinfo{person}{Michael~A.
  Fetterman}, {and} \bibinfo{person}{Peter~M. Chen}.}
  \bibinfo{year}{2008}\natexlab{}.
\newblock \showarticletitle{Execution replay of multiprocessor virtual
  machines}. In \bibinfo{booktitle}{\emph{Proceedings of the 4th International
  Conference on Virtual Execution Environments}}. \bibinfo{pages}{121--130}.
\newblock
\urldef\tempurl%
\url{https://doi.org/10.1145/1346256.1346273}
\showDOI{\tempurl}


\bibitem[\protect\citeauthoryear{Gomez, Neamtiu, Azim, and Millstein}{Gomez
  et~al\mbox{.}}{2013}]%
        {reran}
\bibfield{author}{\bibinfo{person}{Lorenzo Gomez}, \bibinfo{person}{Iulian
  Neamtiu}, \bibinfo{person}{Tanzirul Azim}, {and} \bibinfo{person}{Todd~D.
  Millstein}.} \bibinfo{year}{2013}\natexlab{}.
\newblock \showarticletitle{{RERAN:} timing- and touch-sensitive record and
  replay for Android}. In \bibinfo{booktitle}{\emph{Proceedings of the 35th
  International Conference on Software Engineering}}. \bibinfo{pages}{72--81}.
\newblock
\urldef\tempurl%
\url{https://doi.org/10.1109/ICSE.2013.6606553}
\showDOI{\tempurl}


\bibitem[\protect\citeauthoryear{Google}{Google}{2018}]%
        {octane}
\bibfield{author}{\bibinfo{person}{Google}.} \bibinfo{year}{2018}\natexlab{}.
\newblock \bibinfo{title}{Octane}.
\newblock \bibinfo{howpublished}{\url{https://developers.google.com/octane/}}.
\newblock
\newblock
\shownote{[Online; accessed 16-April-2018].}


\bibitem[\protect\citeauthoryear{Hickson}{Hickson}{2015}]%
        {domStorageSpec}
\bibfield{author}{\bibinfo{person}{Ian Hickson}.}
  \bibinfo{year}{2015}\natexlab{}.
\newblock \bibinfo{title}{{Web Storage (Second Edition)}}.
\newblock \bibinfo{howpublished}{\url{http://www.w3.org/TR/webstorage/}}.
\newblock
\newblock
\shownote{[Online; accessed 30-April-2017].}


\bibitem[\protect\citeauthoryear{Hors, H\'{e}garet, Wood, Nicol, Robie,
  Champion, and Byrne}{Hors et~al\mbox{.}}{2004}]%
        {domSpec}
\bibfield{author}{\bibinfo{person}{Arnaud~Le Hors},
  \bibinfo{person}{Philippe~Le H\'{e}garet}, \bibinfo{person}{Lauren Wood},
  \bibinfo{person}{Gavin Nicol}, \bibinfo{person}{Jonathan Robie},
  \bibinfo{person}{Mike Champion}, {and} \bibinfo{person}{Steve Byrne}.}
  \bibinfo{year}{2004}\natexlab{}.
\newblock \bibinfo{title}{{Document Object Model (DOM) Level 3 Core
  Specification}}.
\newblock
  \bibinfo{howpublished}{\url{http://www.w3.org/TR/2004/REC-DOM-Level-3-Core-20040407/}}.
\newblock
\newblock
\shownote{[Online; accessed 30-April-2017].}


\bibitem[\protect\citeauthoryear{Hu, Azim, and Neamtiu}{Hu
  et~al\mbox{.}}{2015}]%
        {valera}
\bibfield{author}{\bibinfo{person}{Yongjian Hu}, \bibinfo{person}{Tanzirul
  Azim}, {and} \bibinfo{person}{Iulian Neamtiu}.}
  \bibinfo{year}{2015}\natexlab{}.
\newblock \showarticletitle{Versatile yet lightweight record-and-replay for
  Android}. In \bibinfo{booktitle}{\emph{Proceedings of the 2015 Conference on
  Object-Oriented Programming, Systems, Languages, and Applications}}.
  \bibinfo{pages}{349--366}.
\newblock
\urldef\tempurl%
\url{https://doi.org/10.1145/2814270.2814320}
\showDOI{\tempurl}


\bibitem[\protect\citeauthoryear{{Intel Corporation}}{{Intel
  Corporation}}{2015}]%
        {intelSDM}
\bibfield{author}{\bibinfo{person}{{Intel Corporation}}.}
  \bibinfo{year}{2015}\natexlab{}.
\newblock \bibinfo{title}{{Intel 64 and IA-32 Architectures Software
  Developer's Manual V3}}.
\newblock
  \bibinfo{howpublished}{\url{http://www.intel.com/content/www/us/en/architecture-and-technology/64-ia-32-architectures-software-developer-system-programming-manual-325384.html}}.
\newblock
\newblock
\shownote{[Online; accessed 30-April-2017].}


\bibitem[\protect\citeauthoryear{Jin}{Jin}{2015}]%
        {colorgame}
\bibfield{author}{\bibinfo{person}{Linghua Jin}.}
  \bibinfo{year}{2015}\natexlab{}.
\newblock \bibinfo{title}{{linghuaj/Angular-ColorGame}}.
\newblock
  \bibinfo{howpublished}{\url{https://github.com/linghuaj/Angular-ColorGame/tree/angular1}}.
\newblock
\newblock
\shownote{[Online; accessed 16-April-2018].}


\bibitem[\protect\citeauthoryear{King, Dunlap, and Chen}{King
  et~al\mbox{.}}{2005}]%
        {kingVM}
\bibfield{author}{\bibinfo{person}{Samuel~T. King}, \bibinfo{person}{George~W.
  Dunlap}, {and} \bibinfo{person}{Peter~M. Chen}.}
  \bibinfo{year}{2005}\natexlab{}.
\newblock \showarticletitle{{Debugging Operating Systems with Time-Traveling
  Virtual Machines}}. In \bibinfo{booktitle}{\emph{Proceedings of the 2005
  {USENIX} Annual Technical Conference}}. \bibinfo{pages}{1--15}.
\newblock
\urldef\tempurl%
\url{http://www.usenix.org/events/usenix05/tech/general/king.html}
\showURL{%
\tempurl}


\bibitem[\protect\citeauthoryear{Koju, Takada, and Doi}{Koju
  et~al\mbox{.}}{2005}]%
        {koju05}
\bibfield{author}{\bibinfo{person}{Toshihiko Koju}, \bibinfo{person}{Shingo
  Takada}, {and} \bibinfo{person}{Norihisa Doi}.}
  \bibinfo{year}{2005}\natexlab{}.
\newblock \showarticletitle{An efficient and generic reversible debugger using
  the virtual machine based approach}. In \bibinfo{booktitle}{\emph{Proceedings
  of the 1st International Conference on Virtual Execution Environments}}.
  \bibinfo{pages}{79--88}.
\newblock
\urldef\tempurl%
\url{https://doi.org/10.1145/1064979.1064992}
\showDOI{\tempurl}


\bibitem[\protect\citeauthoryear{Lefebvre, Cully, Head, Spear, Hutchinson,
  Feeley, and Warfield}{Lefebvre et~al\mbox{.}}{2012}]%
        {tralfamadore}
\bibfield{author}{\bibinfo{person}{Geoffrey Lefebvre}, \bibinfo{person}{Brendan
  Cully}, \bibinfo{person}{Christopher Head}, \bibinfo{person}{Mark Spear},
  \bibinfo{person}{Norman~C. Hutchinson}, \bibinfo{person}{Mike Feeley}, {and}
  \bibinfo{person}{Andrew Warfield}.} \bibinfo{year}{2012}\natexlab{}.
\newblock \showarticletitle{Execution mining}. In
  \bibinfo{booktitle}{\emph{Proceedings of the 8th International Conference on
  Virtual Execution Environments}}. \bibinfo{pages}{145--158}.
\newblock
\urldef\tempurl%
\url{https://doi.org/10.1145/2151024.2151044}
\showDOI{\tempurl}


\bibitem[\protect\citeauthoryear{Lewis}{Lewis}{2003}]%
        {odb}
\bibfield{author}{\bibinfo{person}{Bill Lewis}.}
  \bibinfo{year}{2003}\natexlab{}.
\newblock \showarticletitle{Debugging Backwards in Time}. In
  \bibinfo{booktitle}{\emph{Proceedings of the Fifth International Workshop on
  Automated Debugging}}.
\newblock


\bibitem[\protect\citeauthoryear{McCandless}{McCandless}{2013}]%
        {browsers-loc}
\bibfield{author}{\bibinfo{person}{David McCandless}.}
  \bibinfo{year}{2013}\natexlab{}.
\newblock \bibinfo{title}{Millions of Lines of Code - Information is
  Beautiful}.
\newblock
  \bibinfo{howpublished}{\url{https://informationisbeautiful.net/visualizations/million-lines-of-code/}}.
\newblock
\newblock
\shownote{[Online; accessed 10-April-2018].}


\bibitem[\protect\citeauthoryear{Mickens, Elson, and Howell}{Mickens
  et~al\mbox{.}}{2010}]%
        {mugshot}
\bibfield{author}{\bibinfo{person}{James Mickens}, \bibinfo{person}{Jeremy
  Elson}, {and} \bibinfo{person}{Jon Howell}.} \bibinfo{year}{2010}\natexlab{}.
\newblock \showarticletitle{Mugshot: {D}eterministic Capture and Replay for
  {J}avaScript Applications}. In \bibinfo{booktitle}{\emph{Proceedings of the
  7th Symposium on Networked Systems Design and Implementation}}.
  \bibinfo{pages}{159--174}.
\newblock


\bibitem[\protect\citeauthoryear{{Microsoft}}{{Microsoft}}{2018}]%
        {clrPM}
\bibfield{author}{\bibinfo{person}{{Microsoft}}.}
  \bibinfo{year}{2018}\natexlab{}.
\newblock \bibinfo{title}{{PerformanceCounter Class}}.
\newblock
  \bibinfo{howpublished}{\url{https://msdn.microsoft.com/en-us/library/system.diagnostics.performancecounter(v=vs.110).aspx}}.
\newblock
\newblock
\shownote{[Online; accessed 16-April-2018].}


\bibitem[\protect\citeauthoryear{{Microsoft Corporation}}{{Microsoft
  Corporation}}{2018}]%
        {chakra-ttd}
\bibfield{author}{\bibinfo{person}{{Microsoft Corporation}}.}
  \bibinfo{year}{2018}\natexlab{}.
\newblock \bibinfo{title}{{Time Travel Debugging with Node-ChakraCore}}.
\newblock \bibinfo{howpublished}{\url{https://aka.ms/NodeTTD}}.
\newblock


\bibitem[\protect\citeauthoryear{{Microsoft Inc.}}{{Microsoft Inc.}}{2017}]%
        {edge-debugger}
\bibfield{author}{\bibinfo{person}{{Microsoft Inc.}}}
  \bibinfo{year}{2017}\natexlab{}.
\newblock \bibinfo{title}{{Microsoft Edge Devtools - Debugger}}.
\newblock
  \bibinfo{howpublished}{\url{https://docs.microsoft.com/en-us/microsoft-edge/devtools-guide/debugger}}.
\newblock
\newblock
\shownote{[Online; accessed 10-April-2018].}


\bibitem[\protect\citeauthoryear{Mozilla}{Mozilla}{2016}]%
        {gecko}
\bibfield{author}{\bibinfo{person}{Mozilla}.} \bibinfo{year}{2016}\natexlab{}.
\newblock \bibinfo{title}{Gecko - MDN}.
\newblock
  \bibinfo{howpublished}{\url{https://developer.mozilla.org/en-US/docs/Mozilla/Gecko}}.
\newblock
\newblock
\shownote{[Online; accessed 10-April-2018].}


\bibitem[\protect\citeauthoryear{{Mozilla}}{{Mozilla}}{2017}]%
        {firefox-debugger}
\bibfield{author}{\bibinfo{person}{{Mozilla}}.}
  \bibinfo{year}{2017}\natexlab{}.
\newblock \bibinfo{title}{{Debugger - Firefox Developer Tools}}.
\newblock
  \bibinfo{howpublished}{\url{https://developer.mozilla.org/en-US/docs/Tools/Debugger}}.
\newblock
\newblock
\shownote{[Online; accessed 10-April-2018].}


\bibitem[\protect\citeauthoryear{O'Callahan, Jones, Froyd, Huey, Noll, and
  Partush}{O'Callahan et~al\mbox{.}}{2017}]%
        {rrtool}
\bibfield{author}{\bibinfo{person}{Robert O'Callahan}, \bibinfo{person}{Chris
  Jones}, \bibinfo{person}{Nathan Froyd}, \bibinfo{person}{Kyle Huey},
  \bibinfo{person}{Albert Noll}, {and} \bibinfo{person}{Nimrod Partush}.}
  \bibinfo{year}{2017}\natexlab{}.
\newblock \showarticletitle{Engineering Record And Replay For Deployability:
  Extended Technical Report}.
\newblock \bibinfo{journal}{\emph{{Computing Research Repository}}}
  \bibinfo{volume}{abs/1705.05937} (\bibinfo{year}{2017}).
\newblock
\showeprint[arxiv]{1705.05937}
\urldef\tempurl%
\url{http://arxiv.org/abs/1705.05937}
\showURL{%
\tempurl}


\bibitem[\protect\citeauthoryear{{Ocariza Jr.}, Bajaj, Pattabiraman, and
  Mesbah}{{Ocariza Jr.} et~al\mbox{.}}{2013}]%
        {frolin-bugs}
\bibfield{author}{\bibinfo{person}{Frolin~S. {Ocariza Jr.}},
  \bibinfo{person}{Kartik Bajaj}, \bibinfo{person}{Karthik Pattabiraman}, {and}
  \bibinfo{person}{Ali Mesbah}.} \bibinfo{year}{2013}\natexlab{}.
\newblock \showarticletitle{An Empirical Study of Client-Side JavaScript Bugs}.
  In \bibinfo{booktitle}{\emph{Proceedings of the 2013 Symposium on Empirical
  Software Engineering and Measurement}}. \bibinfo{pages}{55--64}.
\newblock
\urldef\tempurl%
\url{https://doi.org/10.1109/ESEM.2013.18}
\showDOI{\tempurl}


\bibitem[\protect\citeauthoryear{Pflug}{Pflug}{2015}]%
        {edgehtml}
\bibfield{author}{\bibinfo{person}{Kyle Pflug}.}
  \bibinfo{year}{2015}\natexlab{}.
\newblock \bibinfo{title}{{Introducing EdgeHTML 13, our first platform update
  for Microsoft Edge}}.
\newblock
  \bibinfo{howpublished}{\url{https://blogs.windows.com/msedgedev/2015/11/16/introducing-edgehtml-13-our-first-platform-update-for-microsoft-edge}}.
\newblock
\newblock
\shownote{[Online; accessed 10-April-2018].}


\bibitem[\protect\citeauthoryear{Pothier and Tanter}{Pothier and
  Tanter}{2009}]%
        {bttf}
\bibfield{author}{\bibinfo{person}{Guillaume Pothier} {and}
  \bibinfo{person}{{\'{E}}ric Tanter}.} \bibinfo{year}{2009}\natexlab{}.
\newblock \showarticletitle{Back to the Future: Omniscient Debugging}.
\newblock \bibinfo{journal}{\emph{{IEEE} Software}} \bibinfo{volume}{26},
  \bibinfo{number}{6} (\bibinfo{year}{2009}), \bibinfo{pages}{78--85}.
\newblock
\urldef\tempurl%
\url{https://doi.org/10.1109/MS.2009.169}
\showDOI{\tempurl}


\bibitem[\protect\citeauthoryear{Pothier, Tanter, and Piquer}{Pothier
  et~al\mbox{.}}{2007a}]%
        {omniscientTOD}
\bibfield{author}{\bibinfo{person}{Guillaume Pothier},
  \bibinfo{person}{{\'{E}}ric Tanter}, {and} \bibinfo{person}{Jos{\'{e}}~M.
  Piquer}.} \bibinfo{year}{2007}\natexlab{a}.
\newblock \showarticletitle{Scalable omniscient debugging}. In
  \bibinfo{booktitle}{\emph{Proceedings of the 22nd Annual Conference on
  Object-Oriented Programming, Systems, Languages, and Applications}}.
  \bibinfo{pages}{535--552}.
\newblock
\urldef\tempurl%
\url{https://doi.org/10.1145/1297027.1297067}
\showDOI{\tempurl}


\bibitem[\protect\citeauthoryear{Pothier, Tanter, and Piquer}{Pothier
  et~al\mbox{.}}{2007b}]%
        {pothier-sod}
\bibfield{author}{\bibinfo{person}{Guillaume Pothier},
  \bibinfo{person}{{\'{E}}ric Tanter}, {and} \bibinfo{person}{Jos{\'{e}}~M.
  Piquer}.} \bibinfo{year}{2007}\natexlab{b}.
\newblock \showarticletitle{Scalable omniscient debugging}. In
  \bibinfo{booktitle}{\emph{Proceedings of the 22nd Annual {ACM} {SIGPLAN}
  Conference on Object-Oriented Programming, Systems, Languages, and
  Applications}}. \bibinfo{pages}{535--552}.
\newblock
\urldef\tempurl%
\url{https://doi.org/10.1145/1297027.1297067}
\showDOI{\tempurl}


\bibitem[\protect\citeauthoryear{Ratanaworabhan, Livshits, and
  Zorn}{Ratanaworabhan et~al\mbox{.}}{2010}]%
        {jsmeter}
\bibfield{author}{\bibinfo{person}{Paruj Ratanaworabhan},
  \bibinfo{person}{Benjamin Livshits}, {and} \bibinfo{person}{Benjamin~G.
  Zorn}.} \bibinfo{year}{2010}\natexlab{}.
\newblock \showarticletitle{{JSMeter: Comparing the Behavior of JavaScript
  Benchmarks with Real Web Applications}}. In
  \bibinfo{booktitle}{\emph{{USENIX} Conference on Web Application
  Development}}.
\newblock
\urldef\tempurl%
\url{https://www.usenix.org/conference/webapps-10/jsmeter-comparing-behavior-javascript-benchmarks-real-web-applications}
\showURL{%
\tempurl}


\bibitem[\protect\citeauthoryear{Sen, Kalasapur, Brutch, and Gibbs}{Sen
  et~al\mbox{.}}{2013}]%
        {jalangi}
\bibfield{author}{\bibinfo{person}{Koushik Sen}, \bibinfo{person}{Swaroop
  Kalasapur}, \bibinfo{person}{Tasneem~G. Brutch}, {and} \bibinfo{person}{Simon
  Gibbs}.} \bibinfo{year}{2013}\natexlab{}.
\newblock \showarticletitle{{Jalangi: A selective record-replay and dynamic
  analysis framework for JavaScript}}. In \bibinfo{booktitle}{\emph{Joint
  Meeting of the European Software Engineering Conference and the {ACM}
  {SIGSOFT} Symposium on the Foundations of Software Engineering}}.
  \bibinfo{pages}{488--498}.
\newblock
\urldef\tempurl%
\url{https://doi.org/10.1145/2491411.2491447}
\showDOI{\tempurl}


\bibitem[\protect\citeauthoryear{Stroop}{Stroop}{1935}]%
        {stroop}
\bibfield{author}{\bibinfo{person}{John~Ridley Stroop}.}
  \bibinfo{year}{1935}\natexlab{}.
\newblock \showarticletitle{Studies of interference in serial verbal
  reactions}.
\newblock \bibinfo{journal}{\emph{Journal of Experimental Psychology}}
  \bibinfo{volume}{18}, \bibinfo{number}{6} (\bibinfo{year}{1935}).
\newblock


\bibitem[\protect\citeauthoryear{Undo}{Undo}{2018}]%
        {undodb}
\bibfield{author}{\bibinfo{person}{Undo}.} \bibinfo{year}{2018}\natexlab{}.
\newblock \bibinfo{title}{Reversible Debugging Tools for C/C++ on Linux and
  Android}.
\newblock \bibinfo{howpublished}{\url{https://undo.io/}}.
\newblock
\newblock
\shownote{[Online; accessed 16-April-2018].}


\bibitem[\protect\citeauthoryear{Xia}{Xia}{2013}]%
        {pacman}
\bibfield{author}{\bibinfo{person}{Bingying Xia}.}
  \bibinfo{year}{2013}\natexlab{}.
\newblock \bibinfo{title}{{bxia/Javascript-Pacman}}.
\newblock
  \bibinfo{howpublished}{\url{https://github.com/bxia/Javascript-Pacman}}.
\newblock
\newblock
\shownote{[Online; accessed 16-April-2018].}


\bibitem[\protect\citeauthoryear{Xu, Malyugin, Sheldon, Venkitachalam, and
  Weissman}{Xu et~al\mbox{.}}{2007}]%
        {retrace}
\bibfield{author}{\bibinfo{person}{Min Xu}, \bibinfo{person}{Vyacheslav
  Malyugin}, \bibinfo{person}{Jeffrey Sheldon}, \bibinfo{person}{Ganesh
  Venkitachalam}, {and} \bibinfo{person}{Boris Weissman}.}
  \bibinfo{year}{2007}\natexlab{}.
\newblock \showarticletitle{Retrace: {C}ollecting execution trace with virtual
  machine deterministic replay}. In \bibinfo{booktitle}{\emph{Workshop on
  Modeling, Benchmarking and Simulation}}.
\newblock


\bibitem[\protect\citeauthoryear{Yang, Yang, Xu, Chen, and Zang}{Yang
  et~al\mbox{.}}{2011}]%
        {order}
\bibfield{author}{\bibinfo{person}{Zhemin Yang}, \bibinfo{person}{Min Yang},
  \bibinfo{person}{Lvcai Xu}, \bibinfo{person}{Haibo Chen}, {and}
  \bibinfo{person}{Binyu Zang}.} \bibinfo{year}{2011}\natexlab{}.
\newblock \showarticletitle{{ORDER:} {Object centRic DEterministic Replay for
  Java}}. In \bibinfo{booktitle}{\emph{Proceedings of the 2011 {USENIX} Annual
  Technical Conference}}.
\newblock
\urldef\tempurl%
\url{https://www.usenix.org/conference/usenixatc11/order-object-centric-deterministic-replay-java}
\showURL{%
\tempurl}


\end{thebibliography}
